\documentclass[aps,prd,amssymb,eqsecnum,showpacs]{revtex4}
\usepackage{graphicx}
\usepackage{psfrag}

\begin{document}
\preprint{AEI-2005-005}

\title{Modeling the Black Hole Excision Problem}

\author{B. Szil\'{a}gyi${}^{1,2}$,
	H-O. Kreiss${}^{2,3}$,
	 and
	J. Winicour${}^{1,2}$}
\affiliation{
${}^{1}$Department of Physics and Astronomy,
University of Pittsburgh, Pittsburgh, Pennsylvania 15260\\
${}^{2}$Albert Einstein Institute, Max Planck Gesellschaft,
Am M\"uhlenberg 1, D-14476 Golm, Germany\\
${}^{3}$NADA, Royal Institute of Technology, 10044 Stockholm, Sweden
}

\begin{abstract}

We analyze the excision strategy for simulating black holes. The problem is
modeled by the propagation of quasi-linear waves in a 1-dimensional spatial
region with timelike outer boundary, spacelike inner boundary and a horizon in
between. Proofs of well-posed evolution and boundary algorithms for a second
differential order treatment of the system are given for the separate pieces
underlying the finite difference problem. These are implemented in a numerical
code which gives accurate long term simulations of the quasi-linear excision
problem. Excitation of long wavelength exponential modes, which are latent in
the problem, are suppressed using conservation laws for the discretized system.
The techniques are designed to apply directly to recent codes for the Einstein
equations based upon the harmonic formulation.

\end{abstract}


\maketitle
\section{Introduction}

Two computational difficulties in simulating the emission of gravitational
waves from black hole interactions are the existence of a singularities inside
the black holes and a consistent treatment of the outer boundary so that the
waveform of the gravitational radiation can be accurately extracted.
Fortunately, the observable waves emanate only from the region outside the
black holes and can be simulated, without loss of physical content, in a domain
from which the singularities have been excised by introducing inner boundaries
inside the black holes. However, the treatment of this domain by Cauchy
evolution has had limited success. We present here a quasi-linear model
excision problem whose mathematical analysis is simple enough to reveal some
underlying computational pitfalls and the means to avoid them. In particular,
we show that an accurate simulation of this model excision problem can
be obtained by (i) blending together two evolution algorithms, one of which is
well-posed near the outer boundary and the other well-posed near the inner
boundary, where a superluminal evolution is employed, and (ii) incorporating
discrete conservation laws which control spurious exponential growth. The
results are aimed at numerical formulations of the Einstein equation as second
order quasi-linear wave equations, such as in recent harmonic approaches
~\cite{Garf,szi03,pret}. 

The  problem can be illustrated by considering the propagation of scalar waves
on the fixed background geometry of a spherically symmetric Schwarzschild black
hole. A choice of coordinates commonly adopted in numerical work for excising
the singular region are ingoing Eddington-Finkelstein coordinates $x^\alpha
=(t,r,\theta,\phi)$, in which the Schwarzschild spacetime metric
$g_{\alpha\beta}$ is given by 
\begin{eqnarray}
  g_{\alpha\beta}dx^\alpha dx^\beta  &=& -(1-\frac{2 M}{r})dt^2
       +\frac{4 M}{r}dtdr + (1+\frac{2 M}{r})dr^2  \nonumber \\
       && + r^2 (d\theta^2+\sin^2\theta d\phi^2).
   \label{eq:inEF}
\end{eqnarray}
Here, the spacetime is foliated by the spacelike Cauchy hypersurfaces
$t=const$, which are used to carry out the numerical evolution. The radial
coordinate extends from the location of the singularity at $r=0$ to the
asymptotic region $r\rightarrow \infty$, to which the radiation propagates. The
black hole is located at $r=2M$. In the excision strategy, the simulation is
carried out in a region $R_1\le r \le R_2$, where the inner boundary
satisfies $0<R_1<2M$ and the outer boundary (introduced for the purpose of a
finite coordinate range) satisfies $R_2>>2M$.

The covariant wave equation for a scalar field $u$ is 
\begin{equation}
  \frac{1}{\sqrt{-g}}\partial_\alpha
        (\sqrt{-g}g^{\alpha\beta}\partial_\beta u)=0,
	\label{eq:wave}
\end{equation}
where $g=\det(g_{\alpha\beta})$ and $g^{\alpha\beta}$ is the inverse metric.
In the above coordinates, it takes the explicit form
\begin{eqnarray}
  -(1+\frac{2 M}{r})\partial_t^2 u &+& \frac{4 M}{r}\partial_t\partial_r u
   +(1-\frac{2 M}{r})\partial_r^2 u +\frac{1}{r^2}(\partial_\theta^2 
     +\frac{1}{\sin^2\theta}\partial_\phi^2) u \nonumber \\
   &=& -\frac{2 M}{r^2}\partial_t u  -\frac{2(r- M)}{r^2}\partial_r u
              -2\cot\theta\partial_\theta u.
\label{eq:swave}
\end{eqnarray}

Although this wave equation is non-singular and can be reduced to symmetric
hyperbolic form in the region $R_1\le r \le R_2$, it has some notable features
which complicate a numerical treatment:
\begin{itemize}

\item The appearance of the mixed
$\partial_t\partial_r$ derivative. This is an effect of the
Eddington-Finkelstein coordinates
in which the $(r,\theta,\phi)=const$ observers are moving relative
to the time slicing. In the standard ADM~\cite{adm} formalism,
this gauge effect is described as a non-vanishing {\em shift},
described by the shift vector $\beta^\alpha =(0,\beta^r,0,0)$ where
\begin{equation}
      \beta^r=\frac{1}{1+\frac{r}{2M}}>0.
\label{eq:shift}
\end{equation}

\item The change in sign of the
coefficient of $\partial_r^2$ in passing across the black hole. This change in
sign does not affect the hyperbolicity of the wave equation but it creates a
horizon at $r=2M$ across which waves cannot pass to the exterior. For $r>2M$, 
the $(r,\theta,\phi)=const$-observers move on timelike worldlines; for $r=2M$,
these worldlines are lightlike; and for $r<2M$, they are spacelike.  Thus,
inside the horizon the $\partial_t$ direction becomes spacelike, i.e. an
evolution based upon grid points with constant $r$ is superluminal inside the
horizon.

\item The energy density on the Cauchy hypersurface $t=const$ associated with a
timelike vector $\xi^\alpha$ is ${\cal E}=\xi^\alpha T_{\alpha\beta} n^\beta$,
where $n_\alpha =-(\partial_\alpha t)/\sqrt{-g^{\mu\nu}(\partial_\mu
t)\partial_\nu t}$ is the unit timelike normal to the Cauchy hypersurface and 
\begin{equation}
         T_{\alpha\beta} = (\partial_\alpha u) \partial_\beta u
                   -\frac{1}{2}g_{\alpha\beta}
         g^{\mu\nu}(\partial_\mu u) \partial_\nu u
\end{equation}
is the energy momentum tensor of the scalar field $u$. In this paper
we use the conserved energy 
associated with the choice $\xi^\alpha =t^\alpha$,
defined by $t^\alpha\partial_\alpha =\partial_t$,
whose flow leaves the
Schwarzschild metric invariant.
However, the corresponding energy density
\begin{equation}
            {\cal E}=\frac{1}{2\sqrt{1+\frac{2M}{r}}}\bigg(
     (1+\frac{2M}{r}) (\partial_t u)^2 +(1-\frac{2M}{r})(\partial_r u)^2
        +\frac{1}{r^2}(\partial_\theta u)^2 
          +\frac{1}{r^2\sin^2\theta}\partial_\phi u)^2 \bigg)  
\end{equation}
is positive-definite and provides a norm for the scalar field only  for $r>2M$,
where $t^\alpha$ is timelike. Inside the horizon, $t^\alpha$ is spacelike and
${\cal E}$ represents momentum rather than energy density. This complicates the
use of the energy method for establishing a stable algorithm in the interior
of the horizon. The non-conserved energy associated with the choice
$\xi^\alpha=n^\alpha$ does provide a global norm but we will not pursue that
direction here. Instead, we use mode analysis to treat the inner region.

\end{itemize} 

These features are further complicated by the consideration of proper
boundary conditions for a numerical treatment. They are not peculiar to a
second differential order treatment. In particular, all these features are
prominent in a model excision problem based upon a first order symmetric
hyperbolic treatment of the linear wave equation (\ref{eq:swave}) carried out
in Cartesian coordinates~\cite{lsulsw}. They are also prominent in another
second order discussion of the excision problem~\cite{calab}, in which the
energy norm associated with the choice $\xi^\alpha=n^\alpha$ is used to
treat the stability of the evolution algorithm (\ref{eq:hlinwave}) which we
adopt for the superluminal case.

We study the numerical simulation of the excision problem by introducing
a model 1-dimensional system which retains all the geometric
features of the full problem. In the domain $0\le x\le 1$, we consider
the quasi-linear wave equation
\begin{eqnarray}
    \partial_\alpha(\sqrt{-g}g^{\alpha\beta}\frac{1}{\Phi}\partial_\beta \Phi) &=&    
   \partial_t (\frac {\sqrt{-g}}{\Phi} g^{t\beta} \partial_\beta \Phi) 
  +\partial_x (\frac {\sqrt{-g}}{\Phi} g^{x\beta} \partial_\beta \Phi)
      \label{eq:gnl} \\
      &=&  -\partial_t(\frac{1}{\Phi}\partial_t \Phi)  
       +\partial_t(\frac{a}{\Phi}\partial_x \Phi) 
        +\partial_x(\frac{a}{\Phi}\partial_t \Phi)
        +\partial_x(\frac {b-a^2}{\Phi} \partial_x \Phi)\nonumber \\
	 &=&0
\label{eq:nlwave}
\end{eqnarray}
where the coefficients are smooth and satisfy $ b>0$ and $ a^2 >b$ for $x=0$
and $ a^2 <b$ for $x=1$. These conditions ensure that the $t$-foliation is
spacelike, that the metric determinant satisfies the Lorentzian requirement
$g<0$, that the outer boundary $x=1$ is timelike and that the inner boundary
$x=0$ is spacelike. The shift vector has component $\beta^x=a$. We also
require, analogous to (\ref{eq:shift}), that $a>0$ at $x=0$, which ensures that
the inner boundary is oriented so that all characteristics leave the domain.
There is an event horizon at $ a^2 =b$ across which waves cannot pass in
the outer direction. For initial data satisfying $\Phi>0$, the Cauchy problem
for this equation is well-posed. 

The non-linearity is introduced into (\ref{eq:nlwave}) in order to model
another computational difficulty related to the full Einstein equations. When
harmonic coordinates are introduced to fix the gauge freedom, the Einstein
equations reduce to quasi-linear equations for the components of the metric
whose principle part is identical to (\ref{eq:gnl}). In these harmonic
coordinates, there exists pure gauge solutions (flat metrics)
\begin{equation}
   ds^2= \Phi(-dt^2+dx^2) + dy^2 +dz^2,
	 \label{eq:shgw}
\end{equation}
where
\begin{equation}
   \Phi=1+f(t-x),
   \label{eq:tphi}
\end{equation}
which describing traveling waves with profile $f>-1$. The simulation of
(\ref{eq:shgw}) is one of the standardized tests proposed in the ``Apples
with Apples'' project~\cite{apples,mex1m}, whose purpose is to compare
different numerical treatments of the Einstein equations. The accuracy of
the simulation is complicated by the existence of exponentially growing
gauge waves~\cite{badh}
\begin{equation}
   ds_\lambda^2= \Phi_\lambda(-dt^2+dx^2) + dy^2 +dz^2,
	 \label{eq:shlgw}
\end{equation}
where
\begin{equation}
   \Phi_\lambda=e^{\lambda t}\big( 1+f(t-x)\big ) ,
\label{eq:tlphi}
\end{equation}
which solve the same equations and cannot be damped by dissipative
techniques. For $\lambda\approx 0$, the Cauchy data for (\ref{eq:shlgw})
approximate the Cauchy data for the test gauge wave (\ref{eq:shgw}). The
excitation of this exponential error mode can dominate the simulation after a
relatively short time. An accurate long term simulation of (\ref{eq:shgw})
requires controlling the excitation of (\ref{eq:shlgw}) by some global
method.  Excellent long term accuracy has been obtained for the gauge wave
test (with periodic boundaries) using a code~\cite{lsugw} based upon a
semi-discrete energy estimate derived at the linearized level by the {\em
summation by parts} method~\cite{krsbp, lsusbp,lsude}. Comparable results for
the gauge wave test can also be obtained using analogs of the semi-discrete
monopole conservation laws applied in Sec.~\ref{sec:qlinexcis} to the
quasi-linear wave equation (\ref{eq:nlwave}). See~\cite{badh} for details.

This quasi-linear wave equation closely models the gauge wave problem.
Solutions of (\ref{eq:nlwave}) can be used to construct gauge waves via
(\ref{eq:shgw}). The traveling waves (\ref{eq:tphi}) and (\ref{eq:tlphi}) are
examples. Thus the accurate long term simulation of solutions to
(\ref{eq:nlwave}) can be complicated by the existence of neighboring
solutions which grow exponentially in time, in precise analogy with the gauge
wave problem.

We analyze the quasi-linear excision problem in terms of its elementary
pieces. A stable algorithm for an initial-boundary value problem must also be
stable for the corresponding linearized problem with frozen coefficients.
This problem is considered in Sec.~\ref{sec:lin}. We consider three separate
regions: (i) behavior of the interior evolution, (ii) behavior at a time-like
boundary and (iii) behavior at a spacelike boundary. Treatment of a time-like
boundary in general relativity introduces yet another computational
difficulty related to the constraints. In the harmonic formulation of the
initial-boundary value problem, each component of the metric satisfies a
quasi-linear wave equation whose principle part is identical to
(\ref{eq:wave}). The only well-posed version of the initial boundary value
problem which is known at present to preserve the harmonic constraints
requires a Dirichlet boundary condition for some metric components and a
Neumann boundary condition for the others~\cite{szi03}. For the purpose of
future application of our results to harmonic evolution in numerical
relativity, we concentrate here on such Dirichlet and Neumann boundary
conditions, although the treatment extends to the general maximally
dissipative boundary condition.

The spatial discretization of the wave equation in second differential order
form has many advantages over first order formulations, as discussed
in~\cite{krsecor,kreissort}. These advantages have been recognized in recent
second order harmonic approaches to numerical relativity
~\cite{Garf,szi03,pret}. For the case of zero shift, a comprehensive treatment
of the wave equation in second order form has been given for both
Dirichlet~\cite{krdir} and Neumann boundary conditions~\cite{krneum}. A major
challenge of the excision problem is not just dealing with a shift but with a
shift that is superluminal. This introduces complications even in the absence
of boundaries. Standard implicit~\cite{alcsch} and explicit~\cite{calab}
evolution algorithms for the second order wave equation with shift are unstable
in the superluminal case. We treat this Cauchy problem in
Sec.~\ref{sec:cauchy}. In Sec.~\ref{sec:tbound}, we give a finite difference
algorithm for the subluminal case which is stable for both Dirichlet and
Neumann boundary conditions. We give a detailed analysis, since there is little
treatment of the second order wave equation with shift in the literature.
Whereas treatment of the Dirichlet condition carries over unchanged, a non-zero
shift forces a revision in the treatment of the Neumann condition. The
algorithm  for the subluminal case (the outer algorithm) is well suited for
treating the outer boundary of the excision problem but becomes unstable inside
the horizon, where the shift is superluminal. In Sec.~\ref{sec:sbound}, we
give an alternative finite difference algorithm (the horizon algorithm) which
is stable in this region and admits a stable spacelike excision boundary.

We establish stability by using the method of lines to reduce the wave
equation to ordinary differential equations in time on a spatial grid and
then applying two standard techniques: mode analysis and semi-discrete energy
estimates supplied by the summation by parts technique~\cite{kreiss}. The
energy method is important because it extends directly to the wave equation
with non-constant coefficients in a region with inner and outer boundaries.
We use the energy associated with the $t$-direction along which the spatial
grid propagates. For the region inside the horizon, where this energy does
not provide a positive-definite norm, we use mode analysis to establish a
stable excision algorithm at the inner spacelike boundary.

In Sec.~\ref{sec:linexcis} we treat the linearized excision problem. This
involves blending together the outer and horizon algorithms obtained in
Sec.~\ref{sec:lin}. The additional ingredient is the necessity of non-constant
coefficients in order to model a domain with spacelike inner boundary and
timelike outer boundary. The homogeneous Dirichlet and Neumann boundary
conditions are constructed so that the  energy of the semi-discrete system is
conserved when the coefficients of the wave operator are time independent.
In addition, the algorithm is constructed so that it conserves a monopole
moment associated with the scalar field in the case of time-dependent
coefficients. 

The algorithm is implemented as a numerical code using a fourth
order Runge-Kutta time integration. Two code tests of the linear excision
problem are presented in Sec.~\ref{sec:ltests}: one where the data describe a
pulse completely inside the horizon; the other where the data consist of an
ingoing wave entering the outer boundary. The tests confirm the accuracy and
stability of the finite difference algorithm.

In Sec.~\ref{sec:qlinexcis}, we treat the excision problem for the
quasi-linear wave equation. Although semi-discrete energy conservation is
not considered in the nonlinear case, we construct the algorithm so that
the scalar monopole moment remains conserved. Studies of the gauge wave
problem~\cite{badh} show that either energy conservation or monopole
conservation are sufficient to suppress the exponential mode
(\ref{eq:tlphi}). This is confirmed in the numerical tests of the
quasi-linear excision problem presented in Sec.~\ref{sec:qltests}.

In Sec.~\ref{sec:disc}, we discuss the application to the harmonic
Einstein equations of the algorithms developed here for the model excision
problem. 

\section{Algorithms for the linear wave equation with constant coefficients}
\label{sec:lin}

We consider the wave equation
\begin{equation}
   u_{tt}-2au_{xt}-(b-a^2) u_{xx}=0 \, \, , \, \,  b>0
   \label{eq:linwave}
\end{equation}
obtained from (\ref{eq:nlwave}) by freezing the coefficients and making the
linear approximation $\Phi\approx 1+ u$.  (Where no confusion arises, we use
the abbreviated notation $\partial_tu=u_t$, etc.)  
We treat three separate problems:

\begin{enumerate}

\item The interior Cauchy problem with both signs of $(b-a^2)$.

\item The initial-boundary value problem in the domain $0\le x <\infty$ with
a timelike boundary at $x=0$, i.e. $(b-a^2)>0$.

\item The initial-boundary value problem in the domain $0\le x <\infty$ 
and a spacelike boundary at $x=0$, i.e. $(b-a^2)<0$.

\end{enumerate}

If we can construct stable difference approximations for these
three problems then the combined difference approximation for the 
full linear excision problem is also stable and the rate of growth
is bounded independently of the step size. The reduction of the
initial boundary value problem to Cauchy and halfplane problems
is discussed in~\cite{sundstrom} and in Ch. 12 of~\cite{kreiss}.

\subsection{The Cauchy problem}
\label{sec:cauchy}

The Cauchy problem for (\ref{eq:linwave}) is well-posed for every fixed value
of $a$ and $b>0$. The   Fourier mode $u={\hat u} e^{i\omega x}$ satisfies
\begin{equation} 
      \hat u_{tt}-2ia\omega \hat u_t -(b-a^2)\omega^2=0,
\end{equation}
with solution $\hat u= e^{\lambda t}$
where
\begin{equation} 
     \lambda= i\omega(a \pm\sqrt{b}).
\end{equation}
Thus $\Re (\lambda)=0$ and there are no exponentially growing modes

On a uniform spatial grid $x_\nu=\nu h$, consider the simplest second-order
accurate finite difference approximation to (\ref{eq:linwave}),
\begin{equation}
   W:=u_{tt}-2aD_0 u_{t}-(b-a^2)D_+D_-u=0, 
   \label{eq:olinwave}
\end{equation}
where $D_0 u_\nu=(u_{\nu+1}-u_{\nu-1})/(2h)$,
$D_+ u_\nu=(u_{\nu+1}-u_\nu)/h$ and $D_- u_\nu=(u_\nu-u_{\nu-1})/h$.
The discrete Fourier mode $u_\nu={\hat u}e^{i \omega\nu h}$ satisfies
\begin{equation} 
      \hat u_{tt}-\frac{2ia\sin(\omega h)}{h}\hat u_t 
      +\frac{4(b-a^2)\sin^2(\omega h/2)}{h^2} \hat u=0,
\end{equation}
with solution $\hat u= e^{\lambda t}$
where
\begin{equation} 
     \lambda= \frac{ia\sin(\omega h)}{h}
         \pm\frac{i}{h}\sqrt{a^2\sin^2(\omega h)+4(b-a^2)\sin^2(\omega h/2)}.
\end{equation}
For $b\ge a^2$ it follows that $\Re (\lambda)=0$ but for $b<a^2$ there
are exponentially growing modes, e.g.
\begin{equation} 
     \lambda= \pm \frac{2}{h}\sqrt{a^2-b}.
\end{equation}
for the shortest wavelength mode $\omega h=\pi$.

In order to remedy this problem, we consider the alternative second order
accurate finite difference approximation 
\begin{equation}
     V:= u_{tt}-2aD_0 u_t - (b D_+D_- -a^2 D_0^2) u =0
      \label{eq:hlinwave}
\end{equation}
for which the discrete Fourier mode satisfies      
\begin{equation} 
      \hat u_{tt}-\frac{2ia\sin(\omega h)}{h}\hat u_t 
    +\bigg( \frac{4b\sin^2(\omega h/2) -a^2\sin^2(\omega h)}{h^2} \bigg)
     \hat u=0,
\end{equation}
with solution $\hat u= e^{\lambda t}$
where
\begin{equation} 
     \lambda= \frac{i}{h}\bigg(a\sin(\omega h)
         \pm 2\sqrt{b}\sin(\omega h/2) \bigg).
\end{equation}
Now $\Re (\lambda)=0$ for all values of the coefficients,
subject to $b>0$.

Unfortunately, because the evolution algorithm (\ref{eq:hlinwave}) involves a
wider stencil than (\ref{eq:olinwave}) it is difficult to analyze its
behavior and design a clean algorithm near a timelike boundary. Accordingly,
we will use the {\em outer algorithm} (\ref{eq:olinwave}) near a timelike
boundary and the {\em horizon algorithm} (\ref{eq:hlinwave}) in a region
extending from the excision boundary to the exterior of the horizon.

\subsection{ Timelike boundary problem}
\label{sec:tbound}

We consider the initial-boundary value problem for (\ref{eq:linwave}) with
$b-a^2>0$ in the domain $0\le x<\infty$. Without restriction we can assume 
$b-a^2=1$. Thus we consider the system 
\begin{equation}
   u_{tt}-2au_{tx}-u_{xx}=0.
   \label{eq:slin}
\end{equation}

We can derive an energy estimate. For real functions, let
\begin{equation}
   (u,v)=\int_0^\infty uv\, dx,\quad \|u\|^2=(u,u)  
\label{eq:cnorm}
\end{equation}
denote the $L_2$-scalar product and norm.  The energy of the system is
\begin{equation}
    E=\frac{1}{2}\int_0^\infty (u_t^2 +u_x^2)dx=\frac{1}{2}(\|u_t\|^2+ \|u_x\|^2).
\end{equation}
Integration by parts gives
\begin{eqnarray}
   \frac{1}{2}\partial_t\|u_t\|^2&=&
           (u_t,u_{xx})+2a(u_t,u_{tx})\nonumber \\
     &=&-{1\over 2}\partial_t\|u_x\|^2- u_t(0,t)u_x(0,t)-
       au^2_t(0,t).
\end{eqnarray}
We obtain energy conservation
\begin{equation}
  E_t=\frac{1}{2} \partial_t(\|u_t\|^2+ \|u_x\|^2)=0 
\label{eq:Etconst}  
\end{equation}
if the boundary condition at $x=0$ is given by the Dirichlet condition
\begin{equation}
        u_t(0,t)=0
	\label{eq:cdir}
\end{equation}	
or the Neumann condition 
\begin{equation}
        u_x(0,t)+au_t(0,t)=0.
	\label{eq:cneum}
\end{equation}
Note that the Neumann condition (\ref{eq:cneum}) involves the normal
derivative $g^{x\alpha}\partial_\alpha=\partial_x+a\partial_t$ in the rest
frame intrinsic to the boundary.

We now discuss the difference approximation
\begin{equation}
  W_\nu := u_{\nu tt}-2aD_0u_{\nu t}-D_+D_-u_\nu =0 ,\quad \nu=0,1,2,\ldots .
\label{eq:Wnuconst}
\end{equation}   
Corresponding to (\ref{eq:cnorm}) we use the discrete scalar product
\begin{equation}
 (w,v)_h={h\over 2} w_0v_0+\sum_{\nu=1}^\infty w_\nu v_\nu h. 
\end{equation} 
Using Lemma 11.1.1 of \cite{kreiss}, one can easily derive the
summation by parts rules
\begin{eqnarray}
     (w,D_0v)_h &=&-(D_0w,v)_h-{1\over 4} (w_1+w_{-1})v_0
      -{1\over 4} (v_1+v_{-1})w_0,\label{eq:wDov} \\
    (w,D_-v)_h &=&-(D_+w,v)_h-{1\over 2} (w_1v_0+w_0v_{-1}).
    \label{eq:wD-v}
\end{eqnarray}
From (\ref{eq:Wnuconst}),
\begin{equation}
      \frac{1}{2}\partial_t \|u_t\|_h^2=(u_t,u_{tt})_h=2a(u_t, D_0u_t)_h+
       (u_t,D_-D_+u)_h.
\end{equation}
Using (\ref{eq:wDov}) with $w,v$ replaced by $u_t$, we obtain
\begin{equation}
    2a(u_t,D_0 u_t)_h=-{a\over 2}(u_{1t}+u_{-1t})u_{0t}.
\end{equation}    
Using (\ref{eq:wD-v}) with $w,v$ replaced by $u_t$ and $D_+u$, respectively,
we obtain
\begin{eqnarray}
   (u_t,D_-D_+u)_h &=&
    -{1\over 2}{\partial\over\partial t}\|D_+u\|^2_h
           -{1\over 2}(u_{1t}D_+u_0+u_{0t}D_-u_0) \nonumber \\
    &=&-{1\over 2}{\partial\over\partial t}\|D_+u\|^2_h
-{h\over 4}{\partial\over\partial t}|D_+u_0|^2-u_{0t}D_0 u_0,
\end{eqnarray}
i.e.
\begin{equation}
 \frac{1}{2}\partial_t\left(\|u_t\|^2_h+\|D_+u\|^2_h+{1\over 2}|D_+u_0|^2h\right)=
     -\left(D_0 u_0+{a\over 2}(u_{1t}+u_{-1t})\right) u_{0t}.
\label{eq:econs}
\end{equation}
Thus we obtain discrete energy conservation if we use the boundary
condition
\begin{equation}
   u_{0t}=0
\end{equation}
or
\begin{equation}
    {\cal N}_0:= D_0 u_0 + {a\over 2}(u_{1t}+u_{-1t})) =0,
\end{equation}
corresponding to second order accurate approximations
to the Dirichlet condition (\ref{eq:cdir}) or the Neumann condition
(\ref{eq:cneum}).

The Neumann condition involves the {\em ghost point} $\nu=-1$,  which we
eliminate by means of (\ref{eq:Wnuconst}) to obtain
\begin{equation}
   -\frac{h}{2}W_0+{\cal N}_0 = -\frac{h}{2}u_{tt0}+D_+ u_0
         +au_{t1} .
	 \label{eq:WpNeum}
\end{equation}
We use (\ref{eq:WpNeum}) to update the boundary point via
\begin{equation}  
       u_{tt0}-\frac{2}{h}\bigg(  D_+ u_0 +a u_{t1} \bigg)=0 .
\label{eq:imneum0}
\end{equation}

In the continuum problem, a homogeneous Neumann boundary condition also
leads to conservation of the monopole quantity
\begin{equation}
  Q= \int_0^\infty (u_t-au_x)dx.
  \label{eq:monint}
\end{equation}
This carries over to the conservation of the semi-discrete monopole quantity
\begin{equation}
   {\cal Q}= h\sum_{\nu =1}^\infty (u_{t\nu} -aD_0 u_\nu)+
     \frac{h}{2}(u_{t0} -aD_+ u_0),
\end{equation}
i.e. ${\cal Q}_t=0$, when (\ref{eq:imneum0}) is satisfied. This
semi-discrete monopole conservation is extended to the non-linear
problem in Sec.~\ref{sec:qlinexcis}, where it has proved effective at
suppressing the long wavelength slowly growing mode (\ref{eq:tlphi}). 

\subsection{ Spacelike boundary problem $0\le x <\infty.$ }
\label{sec:sbound}

We now treat the initial-boundary value problem for (\ref{eq:linwave}) in the
halfspace  $0\le x < \infty$ for the case  $a^2 >b$ with $a>0$,  The condition
$a^2 >b$ implies that the boundary $x=0$ is spacelike and the condition $a>0$
implies that the boundary is oriented so that the characteristics leave the
halfspace. Thus no boundary conditions are necessary in the continuum problem.

Since there is no energy estimate of type (\ref{eq:Etconst}), we use mode
analysis with frozen coefficients to formulate a stable discretization near the
boundary. Consider first the continuum system. We study bounded solutions of the
form 
\begin{equation}
         u=e^{st}\cdot\hat u(x) \, , \, \Re(s)>0.
\end{equation}
Introduction of this ansatz into (\ref{eq:linwave}) gives the ordinary
differential equation
\begin{equation}
    s^2 \hat u-2as\hat u_x-(b-a^2)\hat u_{xx}=0.
\label{eq:smode}
\end{equation}
Thus $u$ has the form 
\begin{equation}
        u=e^{st} (\sigma_1 e^{\mu_1 x}+ \sigma_2 e^{\mu_2 x})	
\end{equation}
where $\mu_j$ are the solutions of the characteristic equation
\begin{equation}
    -s^2+2as\mu+(b-a^2)\mu^2 =0 ,
\label{eq:smodech}    
\end{equation}
i.e.
\begin{equation}
   \mu_{1,2}=-{as\over b-a^2}\pm\sqrt{ {a^2s^2\over (b-a^2)^2}+
   {s^2(b-a^2)\over (b-a^2)^2}}=
    {as\over a^2-b}\pm {s\sqrt{b}\over a^2-b}.
\end{equation}
Since $a>\sqrt{b},$ all roots satisfy $\Re(\mu_{1,2})>0$ for $\Re (s)>0$ and
there are no bounded solutions. Thus the Kreiss condition (see~\cite{kreiss},
Ch. 10) is trivially satisfied and the problem is well posed.

As difference approximation for the halfplane problem we use
the horizon evolution algorithm (\ref{eq:hlinwave}), i.e.
\begin{equation}
   u_{\nu tt}-2a{(u_{\nu+1}-u_{\nu-1})_t\over 2h}
   -b{(u_{\nu+1}-2u_\nu +u_{\nu-1})\over h^2}
   +a^2{(u_{\nu+2}-2u_\nu +u_{\nu-2})\over 4h^2}=0
\label{eq:hlinwave2}  
\end{equation}
for $\nu=2,3,\ldots$. We need two extra boundary conditions
to determine $u_0$ and $u_1$. 
We use the extrapolation conditions
\begin{equation}
   h^3 D_+^3 u_0=0,\quad h^3 D_+^3 u_1=0. 
\label{eq:uextrap}
\end{equation}   

Now we study bounded solutions of type
\begin{equation}
       u_\nu=e^{st}\sum_{j=1}^4 \sigma_j\kappa_j^\nu,\quad
                              \Re (s)>0
\label{eq:extrmode}
\end{equation} 
where $\kappa_j$ are solutions of the characteristic equation 
\begin{equation}
   \tilde s^2-a\tilde s (\kappa -{1\over\kappa})-b(\kappa -2+{1\over\kappa})
+{a^2\over 4} (\kappa -{1\over\kappa})^2=0 , \quad \tilde s=hs.
\label{eq:tilde}
\end{equation}

\medskip

\noindent {\bf Lemma}
\medskip

(i) (\ref{eq:tilde}) has no solutions with
\begin{equation}
   |\kappa|=1\quad {\rm for} \, \Re(\tilde s) >0. 
\end{equation}

(ii) There are exactly two solutions $\kappa_1,\kappa_2$ with
\begin{equation}
   |\kappa_j|<1,~j=1,2 \quad {\rm for} \,  \Re(\tilde s) >0.
\end{equation}

(iii) For $\tilde s\to 0$,
\begin{equation}
   \lim_{\tilde s\to 0} \kappa_1= -(1-{2b\over a^2})+
         \sqrt{(1-{2b\over a^2})^2-1},\quad
   \lim_{\tilde s\to 0} \kappa_2= -(1-{2b\over a^2})-
          \sqrt{(1-{2b\over a^2})^2-1}.
\label{eq:zerlim}
\end{equation}

\medskip\noindent
{\it Proof.} Part (i) follows from the stability of the Cauchy problem,
established in Sec.~\ref{sec:cauchy}.
\medskip
For $\tilde s$ real, as $\tilde s\to\infty$ the solutions of (\ref{eq:tilde}) with
$|\kappa|<1$  converge to the solutions of
\begin{equation}
  \tilde s^2+{a^2\over 4}\,{1\over\kappa^2}=0,\quad
    {\rm i.e.}\quad \kappa_{1,2} =\pm{ia\over 2}\,{1\over\tilde s}.
\end{equation}
Correspondingly, the solutions of (\ref{eq:tilde}) with $|\kappa|>1$ 
converge to the solutions of
\begin{equation}
     \tilde s^2+{a^2\over 4}\kappa^2=0,\quad
         {\rm i.e.}\quad \kappa_{3,4}=\pm {2\tilde s\over ia}.
\end{equation}
Since the four roots $\kappa(\tilde s)$ of (\ref{eq:tilde}) are continuous
functions of $\tilde s$ and $|\kappa|\ne 1$ for $\Re(\tilde s)>0,$ there are
exactly two roots $\kappa_{1,2}$ with $|\kappa_{1,2}|<1$ for all $\tilde s$
with $\Re(\tilde s) >0$, thus establishing (ii).
\medskip
For $\tilde s=0,$ we have
\begin{equation}
    -b{(\kappa-1)^2\over \kappa}+{a^2\over 4}{(\kappa^2-1)^2\over \kappa^2}=
    {(\kappa-1)^2\over \kappa}\left(
      -b+{a^2\over 4}{(\kappa+1)^2\over \kappa}\right)=0,
\end{equation}
i.e.
\begin{equation}
    (\kappa-1)^2=0\quad {\rm or}\quad
    (\kappa+1)^2-{4b\over a^2}\kappa=
     \kappa^2+2(1-{2b\over a^2})\kappa +1=0.
\end{equation}
Therefore,
\begin{equation}
   \kappa_{1,2}=-(1-{2b\over a^2})\pm
   \sqrt{(1-{2b\over a^2})^2 -1},\quad \kappa_{3,4}=1.
\end{equation}
A simple perturbation analysis shows that 
$ |\kappa_{3,4}|\sim |e^{\mu_{1,2}h}|>1$ for
$|\tilde s|<\!< 1,$ $\Re(\tilde s)>0$,  i.e.
$\kappa_{3,4}$ correspond to the solution of (\ref{eq:smodech}). Since
$a^2>b,$ we have $|\kappa_{1,2}|= 1$ for $\tilde s\to 0$ and (\ref{eq:zerlim})
follows. This proves the lemma.

The lemma shows that $\sigma_3=\sigma_4=0.$ To account for possible double
roots we write (\ref{eq:extrmode}) as
\begin{equation}
    u_\nu=e^{st}(\sigma_1\kappa_1^\nu+
       \sigma_2{\kappa_2^\nu-\kappa_1^\nu\over
         \kappa_2-\kappa_1})\quad {\rm if}~ \kappa_1 \ne \kappa_2, 
\label{eq:doub}
\end{equation}
which becomes
\begin{equation}
     u_\nu=e^{st}(\sigma_1\kappa_1^\nu+\sigma_2\nu\kappa_2^{\nu-1})
      \quad {\rm if}~ \kappa_1 =\kappa_2.
\end{equation}
Here $\sigma_1,\sigma_2$ are determined by the boundary conditions (\ref{eq:uextrap}), i.e.
\begin{eqnarray}
     \sigma_1(\kappa_1-1)^3+{\sigma_2\over \kappa_2-\kappa_1}
        \left((\kappa_2-1)^3-(\kappa_1-1)^3\right)&=0&,
              \nonumber \\
     \sigma_1\kappa_1(\kappa_1-1)^3+{\sigma_2\over \kappa_2-\kappa_1}
        \left(\kappa_2(\kappa_2-1)^3-\kappa_1(\kappa_1-1)^3\right)&=0&.
\label{eq:sigma}
\end{eqnarray}
The determinant of this system is
\begin{equation}
     {\rm Det}= (\kappa_1-1)^3 (\kappa_2-1)^3. 
\label{eq:det} 
\end{equation}

We can now prove:

\medskip

\noindent {\bf Theorem}.  (\ref{eq:doub}), (\ref{eq:sigma}) has only the
trivial solution $(\sigma_1=\sigma_2=0)$ for $\Re(s)\ge 0$. Therefore the
difference approximation (\ref{eq:hlinwave2}), (\ref{eq:uextrap}) is stable.
\par
\medskip\noindent
{\it Proof.} By (\ref{eq:det}), we have $\kappa_1=1$ or $\kappa_2=1.$ Then, by
(\ref{eq:tilde}), $s=0$ and, by (\ref{eq:zerlim}), $\kappa_1\ne 1,~\kappa_2\ne
1,$ which is a contradiction. Therefore the Kreiss condition is satisfied. In
Ch. 12 of~\cite{kreiss},  it is proved that the Kreiss condition implies
stability.

\medskip

\noindent{\bf Remark 1.} The Ryabenkii-Godonov condition states that there are
no bounded solutions of type (\ref{eq:doub}) for ${\cal R}(s)>0.$ The Kreiss
condition requires that there are no bounded solutions for ${\cal R}(s)\ge 0.$
The  Ryabenkii-Godonov condition is a necessary condition while the Kreiss
condition is a sufficient condition for stability. (See Ch. 12
of~\cite{kreiss}.)

\medskip

\noindent {\bf Remark 2.} Mode analysis is closely connected with the Laplace
transform.  Therefore the stability estimates are expressed in terms of
$\int_0^T \|u(\cdot,t)\|^2_h dt.$ As explained in Ch. 12 of~\cite{kreiss}, one
can use these estimates to obtain standard energy estimates.

\medskip

\noindent {\bf Remark 3.} The stability theory in~\cite{sundstrom}
and in Ch. 12 of~\cite{kreiss} is
only developed for first order systems. However, it generalizes directly to
second order equations. Also in~\cite{sundstrom}, it is indicated how to prove
that the Principle of Frozen Coefficients holds. The approximation with
smooth coefficients is stable if the system is strictly hyperbolic and
the Kreiss condition holds. 

\subsection{Time discretization}

The results in Sec's~(\ref{sec:tbound}) and (\ref{sec:sbound}) prove that the
outer algorithm gives a stable difference approximation for the halfplane
problem with timelike boundary and that the horizon algorithm gives a stable
difference approximation for the halfplane problem with a spacelike boundary.
In Sec.~(\ref{sec:linexcis}), these results will be combined to give a stable
difference approximation for the strip problem with spacelike inner boundary
and timelike outer boundary.

For the time discretization, we use the method of lines. The
spatial discretization reduces the problem to a 
large system of stable ODE's
\begin{equation}
     {\bf u}_{tt}=\frac{1}{h}{\bf Au}_t+\frac{1}{h^2}{\bf Bu}.
\end{equation}
Introducing 
\begin{equation}
   {\bf u}_t=\frac{1}{h}{\bf v}, 
\end{equation}
we obtain the first order system
\begin{equation}
\left(
\begin{array}{c}
{\bf v} \\
{\bf u}    
\end{array}
\right)_t =  \frac{1}{h}
\left( 
\begin{array}{cc}
{\bf B} & {\bf A} \\
{\bf I} & {\bf 0}   
\end{array}
\right)  
\left( 
\begin{array}{c}
{\bf v} \\
{\bf u}    
\end{array}
\right) . 
\label{eq:fo}
\end{equation}
We solve the system numerically using a 4th order
Runge-Kutta time integrator. The time step is limited by
the CFL condition determined from the Cauchy problem.

\section{The linear excision problem}
\label{sec:linexcis}

We model a pulse of energy propagating into a horizon using the
wave equation 
\begin{equation}
     -\partial_t^2 u +\partial_t(a\partial_x u) +\partial_x(a\partial_t u)
             +\partial_x\bigg((1-a^2)\partial_x u) \bigg) =0
\label{eq:lxwave}
\end{equation}
on the interval $0\le x\le 1$, where we set
\begin{eqnarray}
        a&=& \frac{5}{4} -x \, , \quad 0\le x \le \frac{1}{2} \nonumber \\
        a&=& \alpha(x) \, , \quad \frac{1}{2} \le x \le \frac{3}{4} \nonumber \\
	a&=&\frac{1}{2}  \, , \quad \frac{3}{4} \le x \le 1,
\label{eq:exc_background}
\end{eqnarray}
with $\alpha$ providing a smooth monotonic transition between the
inner and outer regions. The underlying metric is Lorentzian and
(\ref{eq:lxwave}) is hyperbolic. The inner boundary is spacelike, the outer
boundary is timelike and there is an event horizon at $x=1/4$.

The solution to (\ref{eq:lxwave}) in the region $3/4 \le x \le 1$ has the
exact form
\begin{equation}
     u=f(t+\frac{2}{3}x)+g(t-2x)
\label{eq:lsol}
\end{equation}
consisting of an ingoing wave $f$ and an outgoing wave $g$.
Near the inner boundary all waves are ingoing, as can be see by freezing
the coefficient $a$ to obtain the solution
\begin{equation}
     u=f(t+\frac{4}{9}x)+g(t+4x),
\end{equation}
which is valid in the short wavelength limit.

Our goal is to simulate a wave $f$ incident on the boundary at
$x=1$ which propagates across the horizon, while being partially backscattered.
We also check that a pulse of compact support inside the horizon does not
propagate into the exterior region.

\subsection{Linear algorithm with non-constant coefficients}

First we must modify the evolution-boundary algorithm to account for
non-constant coefficients, as required for the excision problem. We
consider the equation
\begin{equation}
     -\partial_t^2 u +\partial_t(a\partial_x u) +\partial_x(a\partial_t u)
             +\partial_x (c\partial_x u) =0,
\label{eq:lcwave}
\end{equation}
where $a=a(t,x)$ and $c=c(t,x)$.  The energy associated with
$t^\alpha\partial_\alpha=\partial_t$ is
\begin{equation}
        E=\frac{1}{2}\int_0^1 ( u_t^2 +c u_x^2 ) dx.
\label{eq:uenergy}
\end{equation}
Note that for $c<0$, $E$ is not necessarily positive. For the case
$a_t=c_t=0$, energy conservation holds in the form
\begin{equation}
      \partial_t E= F|_{x=1}-F|_{x=0}
      \label{eq:encons}
\end{equation}
where the boundary flux is
\begin{equation}
   F=u_t(cu_x+a u_t).
\end{equation}
With this generalization, the homogeneous Dirichlet condition (\ref{eq:cdir})
remains $u_t=0$ but the homogeneous Neumann condition (\ref{eq:cneum}) must
be modified to 
\begin{equation}
  cu_x+a u_t =0.
  \label{eq:nchomneum}
\end{equation}

We modify the outer algorithm to
\begin{equation}
  W:= u_{tt}-\partial_t(aD_0 u)-D_0(a u_t)
       -\frac{1}{2} D_-\bigg( (A_+ c) D_+u \bigg)
         -\frac{1}{2} D_+\bigg( (A_- c) D_- u\bigg) =0,
   \label{eq:onclinwave}
\end{equation}
where
\begin{equation}
    A_\pm f_\nu=\frac {f_{\nu\pm 1} +f_\nu}{2}.
\end{equation}
The last two terms in (\ref{eq:onclinwave}) are equal because of the identity
\begin{equation} 
  D_-\bigg((A_+ f)D_+ g\bigg)= D_+\bigg((A_- f)D_- g\bigg),
\end{equation}
but it is advantageous to express $W$ in a form which is manifestly
reflection invariant.

Consider now semi-discrete energy conservation for the case $a_t=c_t=0$.
We modify the semi-discrete version of the energy to
\begin{equation} 
    E=h\sum_1^{N-1} {\cal E}     
   +\frac{h}{4}\bigg((u_{tN})^2+ (A_- c_N)(D_-u_N)^2+(u_{t0})^2
    +(A_+ c_0)(D_+u_0)^2\bigg)
    \label{eq:ncde}
\end{equation}
where
\begin{equation}
  {\cal E}=\frac{1}{2}u_t^2 +\frac{1}{4}(A_+ c)(D_+u)^2 
                + \frac{1}{4}(A_- c)(D_-u)^2 .  
    \label{eq:ncded}
\end{equation}
It follows that
\begin{eqnarray}
     {\cal E}_t-u_t W &=&  D_0 (a u_t^2)
  -\frac{h}{2}\bigg( (D_+u_t)D_+(au_t)-(D_-u_t)D_- (au_t) \bigg)\nonumber \\
   &+&\frac{1}{2}u_t D_-\bigg((A_+ c)D_+u\bigg)
          +\frac{1}{2}(D_+ u_t) (A_+ c)D_+u \nonumber \\
    &+&\frac{1}{2}u_t D_+\bigg((A_- c)D_-u\bigg)
          +\frac{1}{2}(D_- u_t) (A_- c)D_-u .
\end{eqnarray}
Following the procedure leading to (\ref{eq:econs}), the summation of this
expression  gives the semi-discrete flux conservation law
\begin{eqnarray}
 \partial_t E &=&\frac{1}{2}u_{tN} \bigg (hW_N+ (A_+ c_N)D_+ u_N+(A_- c_N)D_- u_N
  +  (A_+ a_N)u_{t(N+1)}+ (A_- a_N)u_{t(N-1)}\bigg) \nonumber \\  
    &-& \frac{1}{2}u_{t0} \bigg (-hW_0+ (A_+ c_0)D_+ u_0+(A_- c_0)D_- u_0
  +  (A_+ a_0)u_{t1}+ (A_- a_0)u_{t(-1)}\bigg).
\label{eq:ncecons}
\end{eqnarray}  

We express the Neumann condition (\ref{eq:nchomneum}) in the second
order accurate form 
\begin{equation}
{\cal N}':= \frac{1}{2}\bigg( (A_+ c)D_+ u+(A_- c)D_- u
  +  (A_+ a)T_+u_t+ (A_- a)T_- u_t \bigg)=0,
  \label{eq:phomneum}
\end{equation}
where $T_\pm f_\nu=f_{\nu \pm 1}$.
At the boundaries, ${\cal N}'_N$ and ${\cal N}'_0$ contain ghost points
outside the computational domain, which we eliminate via the relations
\begin{eqnarray}
  \frac{h}{2}W_N+{\cal N}'_N &=& \frac{h}{2}u_{ttN}+(A_- c_N)D_- u_N
         +(A_- a_N)u_{t(N-1)} 
	 \label{eq:ncwneumN} \\    
   -\frac{h}{2}W_0+{\cal N}'_0 &=& -\frac{h}{2}u_{tt0}+(A_+ c_0)D_+ u_0.
         +(A_+ a_0)u_{t1} .
	 \label{eq:ncwneum0}
\end{eqnarray}
Then, requiring that the wave equation be satisfied at the boundary
points, i.e. that $W_N=W_0=0$, the conservation law (\ref{eq:ncecons})
implies that the Neumann boundary conditions
\begin{eqnarray}
  {\cal N}'_N:&=&\frac{h}{2}u_{ttN}+(A_- c_N) D_-u_N+
          (A_- a_N)u_{t(N-1)}=0
   \label{eq:ncimneumN}   \\    
  {\cal N}'_0:&=&-\frac{h}{2}u_{tt0}+(A_+ c_0) D_+ u_0
          +(A_+ a_0)u_{t1}=0  
\label{eq:ncimneum0}
\end{eqnarray}
are dissipative.

Energy conservation holds in the continuum theory only when the coefficients of
the wave operator are time independent. However, in the continuum problem,
conservation of the monopole quantity (\ref{eq:monint}) holds for general
time-dependent coefficients in the case of homogeneous Neumann boundary
conditions. This carries over  to the conservation of the semi-discrete
quantity
\begin{equation}
   {\cal Q}= h\sum_{\nu =1}^{N-1}(u_{t\nu} -aD_0 u_\nu)+
     \frac{h}{2}(u_{tN} -aD_- u_N +u_{t0} -aD_+ u_0)
\end{equation}
when the Neumann condition (\ref{eq:phomneum}) has the correction
\begin{equation}
 {\cal N}:= {\cal N}'+\frac{h^2}{4} a_tD_+ D_- u=0,
  \label{eq:homneum}
\end{equation}
which preserves second order accuracy.
As a result of this correction, (\ref{eq:ncimneumN}) and
(\ref{eq:ncimneum0}) are modified to  
\begin{eqnarray}
  \frac{h}{2}W_N+{\cal N}_N:=\frac{h}{2}(u_{ttN}-a_{tN}D_- u_N) +(A_- c_N) D_-u_N+
          (A_- a_N)u_{t(N-1)}=0
   \label{eq:nctimneumN}   \\    
  -\frac{h}{2}W_0+{\cal N}_0:=-\frac{h}{2}(u_{tt0}-a_{t0}D_+ u_0)+(A_+ c_0) D_+ u_0
          +(A_+ a_0)u_{t1}=0 .
\label{eq:nctimneum0}
\end{eqnarray}
Setting $W_\nu=0$ for $1\le\nu\le N-1$, a
straightforward calculation then gives
\begin{equation}
      {\cal Q}_t =\frac{h}{2}W_N+{\cal N}_N +\frac{h}{2}W_0-{\cal N}_0.
      \label{eq:fluxcons}
\end{equation}
Thus, again requiring that $W_N=W_0=0$, the boundary conditions
\begin{eqnarray}
  {\cal N}_N=\frac{h}{2}(u_{ttN}-a_{tN}D_- u_N) +(A_- c_N) D_-u_N+
          (A_- a_N)u_{t(N-1)}=0
   \label{eq:nctlneumN}   \\    
 {\cal N}_0=-\frac{h}{2}(u_{tt0}-a_{t0}D_+ u_0)+(A_+ c_0) D_+ u_0
          +(A_+ a_0)u_{t1}=0 ,
\label{eq:nctlneum0}
\end{eqnarray}
imply ${\cal Q}_t=0$. Thus we have established that ${\cal Q}_t=0$ in the
general case and $E_t=0$ when the coefficients are frozen in time.

Note that when the coefficients are not frozen in time, an energy estimate
still applies provided $a,a_t,c,c_t$ are bounded functions, with $c>0$. For
simplicity, consider the case with periodic boundaries. Then for the continuum
problem (\ref{eq:lcwave}) and (\ref{eq:uenergy}) imply
\begin{equation}
         E_t=\int_0^1 ( a_t u_t u_x +\frac{c_t}{2} u_x^2 ) dx.
\label{eq:et}
\end{equation}
The inequality $2fg\le f^2+g^2$ gives
\begin{equation}
    a_t u_t u_x \le \frac {|a_t|}{2\sqrt{c}}(u_t^2+cu_x^2)\le K_1{\cal E}
\end{equation}
where the constant $K_1$ satisfies $K_1 \ge |a_t|/\sqrt{c}$; and 
\begin{equation}
    \frac{c_t}{2} u_x^2 \le \frac {|c_t|}{2c}(u_t^2+cu_x^2)\le K_2{\cal E},
\end{equation}
where the constant $K_2$ satisfies $K_2 \ge |c_t|/c$. Integration of
(\ref{eq:et}) then gives $E_t\le (K_1+K_2)E$ so that
\begin{equation}
                E\le E_0 e^{(K_1+K_2)t},
\label{eq:eest}
\end{equation}
where $E_0$ is the initial energy. A straightforward calculation based upon the
summation
\begin{equation} 
          E_t=h\sum_1^N ({\cal E}_t -u_t W)   
\end{equation}
leads to a semi-discrete energy estimate analogous to (\ref{eq:eest}).

Finally, in the inner region, we modify the horizon algorithm
(\ref{eq:hlinwave}) to the non-constant coefficient form
\begin{equation}
   V:= u_{tt}-\partial_t(aD_0 u)-D_0(a u_t)
       -\frac{1}{2} D_-\bigg( (A_+ b) D_+u \bigg)
         -\frac{1}{2} D_+\bigg( (A_- b) D_- u\bigg)
        +D_0(a^2 D_0u) =0 , 
   \label{eq:nchlinwave}
\end{equation}
where $c=b-a^2$. 

\subsection{Blending the outer and horizon algorithms}

For the purpose of excision, we need a prescription for switching from the
outer algorithm $W=0$ to the horizon algorithm $V=0$. We introduce a smooth,
monotonic blending function $f(c)=f(b-a^2)$, satisfying $f=1$ for $c\le 0$ and 
$f=0$ for $c \ge 1/2$. Referring to (\ref{eq:onclinwave}) and
(\ref{eq:nchlinwave}), we then  use the blended evolution algorithm $B=0$,
where
\begin{equation}
   B= W - D_-\bigg( (A_+ (f a^2)) D_+ u \bigg)+ D_0\bigg(fa^2 D_0 u\bigg)
     = W +\frac{h^2}{4} D_+ D_- (fa^2D_+ D_- u)  ,
   \label{eq:blend}
\end{equation}
which reduces to $W$ near the outer boundary and to $V$ inside the horizon.

When $\partial_t g^{\alpha\beta}=0$, the blended algorithm satisfies
a semi-discrete energy conservation law. For periodic boundary conditions
(identifying the points $\nu=0$ and $\nu=N$), this takes the form
$\partial_t E_B=0$ where
\begin{equation} 
    E_B=h\sum_1^{N} {\cal E}_B ,   
\end{equation}
with
\begin{equation}
  {\cal E}_B={\cal E} 
	       +\frac{fh^2 a^2}{8}(D_+ D_- u)^2.  
\end{equation}
(Here, as before, ${\cal E}$ is not necessarily positive inside the horizon.)

For the boundary conditions of the excision problem, the energy (\ref{eq:ncde})
is modified in the same way, 
\begin{equation} 
    E_B=E+h\sum_1^{N-1} \frac{fh^2 a^2}{8}(D_+ D_- u)^2 .
    \label{eq:be}
\end{equation}
Since the support of the blending function $f$ is isolated from the outer
timelike boundary, the semi-discrete energy flux through the outer boundary
remains given by the first term in (\ref{eq:ncecons}). 

\subsection{Tests of the linear excision algorithm}
\label{sec:ltests}

In order to validate the linear algorithm a set of test-runs was performed
where the outer boundary condition at $x=1$  was either Dirichlet or Neumann,
while at $x=0$ the extrapolation conditions~(\ref{eq:uextrap}) were applied.
The gridstep was chosen to be $h = 1/(\rho \cdot 2000)$, $\rho = 1, 2, 4$ with
the time-step set to $\Delta t = h / 10 $. The background coefficients were set
according to~(\ref{eq:exc_background}) with $\alpha(x) = 245/4-513\, x+1728\,
x^2-2880\, x^3+2368\, x^4-768\, x^5$ so that $C^2$ differentiability results.

\begin{figure}
\psfrag{xlabel}{$x$ axis}
\psfrag{ylabel}{$u$}
  \includegraphics[width=8cm]{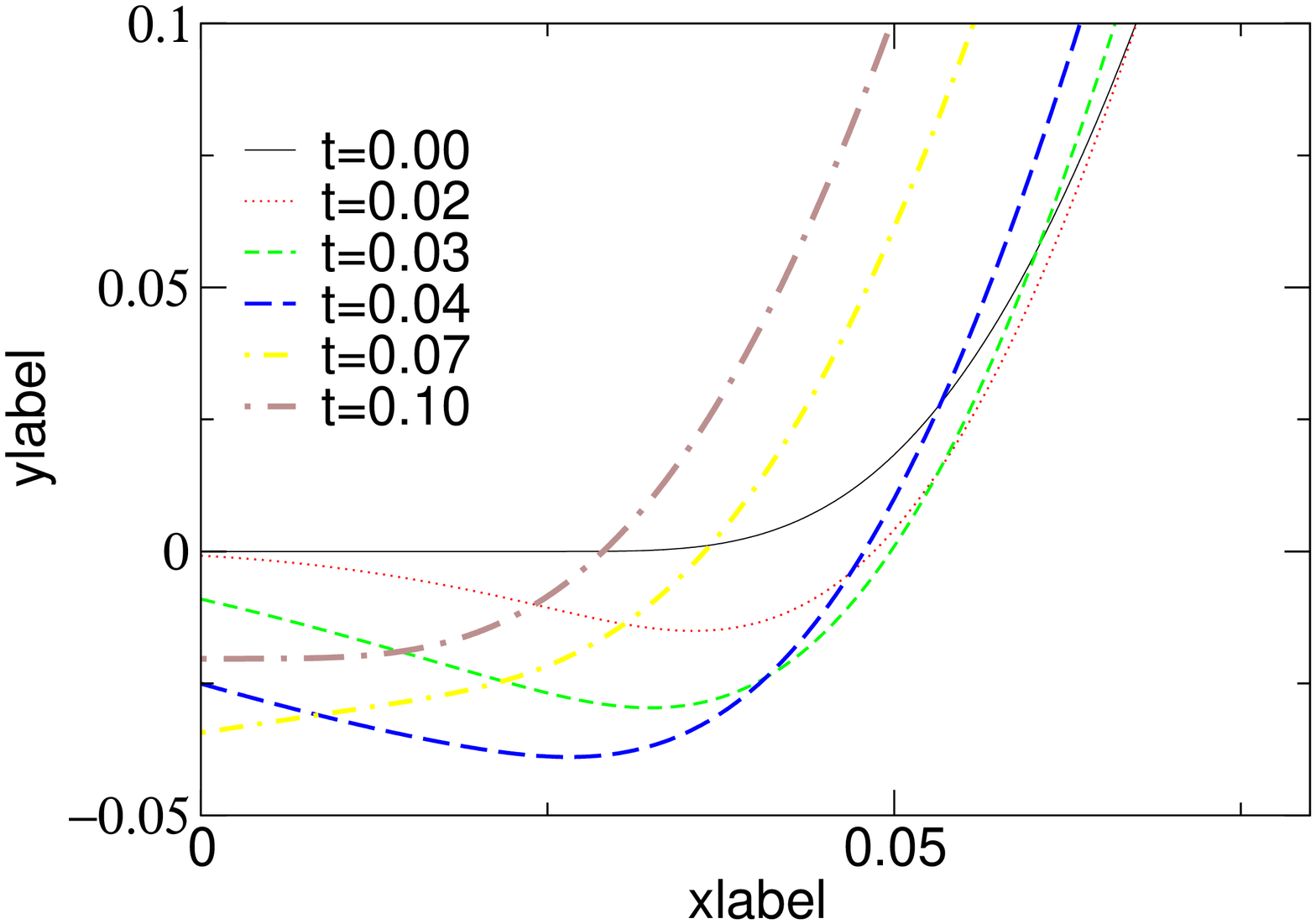}
  \includegraphics[width=8cm]{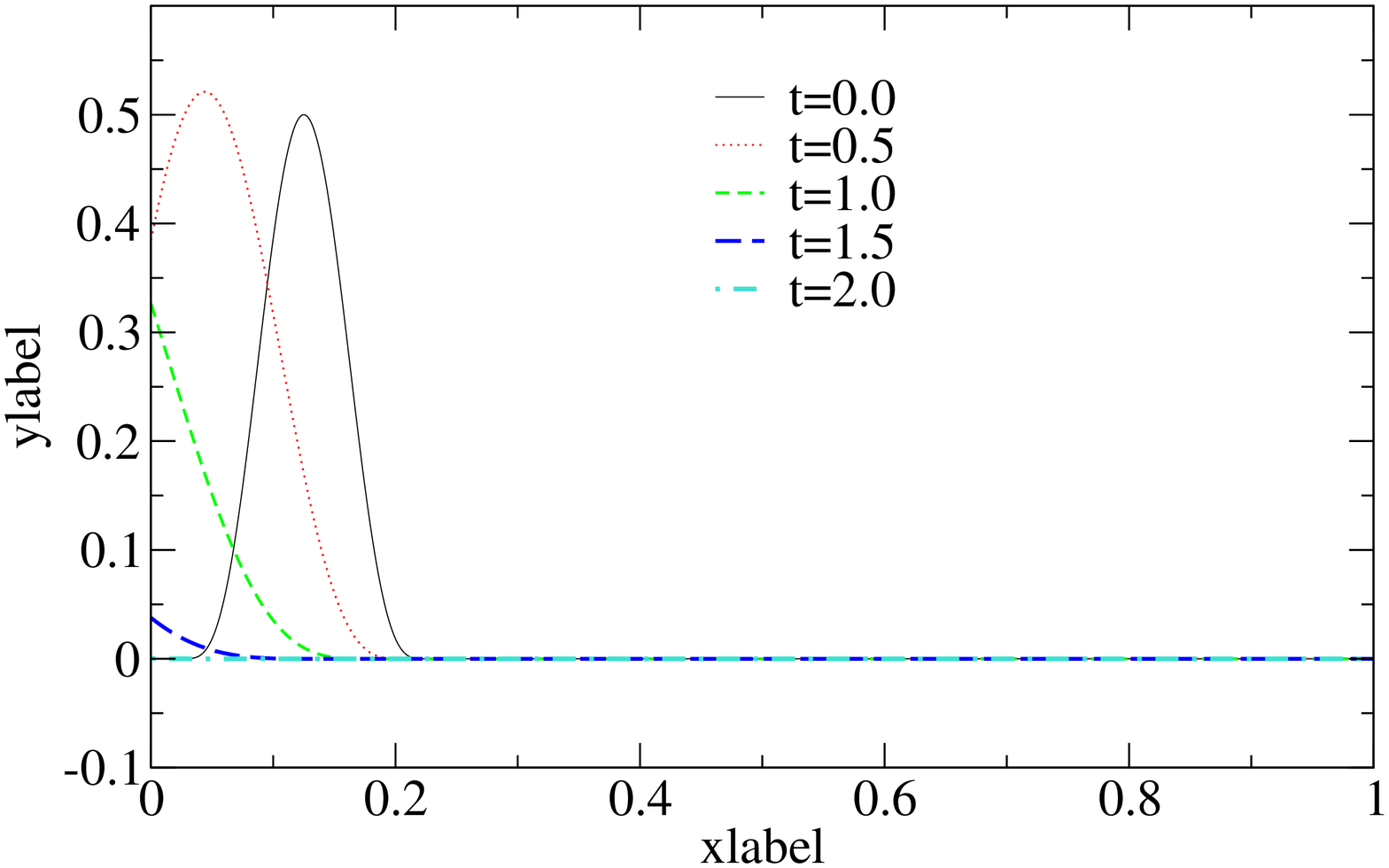}
\caption{Sequence of snapshots showing the initial pulse falling into the
excised region. The left graph shows the early behavior as the faster mode
(with negative amplitude) propagates through the excision boundary. The right
graph shows the slower mode making its way through the excision boundary, with
no observable trace remaining at $t=2.0$.}
\label{fig:lin.in.slides}
\end{figure}

In the first set of runs the initial data at $t=0$ were a $C^3$ pulse of compact
support inside the horizon with $u_t(0,x)=0$ and
\begin{eqnarray}
    u(0,x) &=& 0.5 \times \big[ 4\, \xi\, (1-\xi) \big]^4
     \quad \mbox{for} \quad x_0 \leq x \leq x_0 + 0.2 
     \label{eq:cpulse} \\
         u(0,x) &=& 0  \quad \mbox{elsewhere}, \nonumber
\end{eqnarray} 
with $\xi(\lambda) = ( \lambda - 0.025 ) / 0.2$ . Homogeneous Dirichlet
data $u(t,1)=0$ was given at the outer boundary. At the initial location of
the pulse inside the horizon, both characteristics point to the left. The
condition $u_t(0,x)=0$ implies that the initial data is the superposition
of two pulses, whose amplitudes have opposite signs, which propagate along
these two characteristics.  This is clearly seen in
Fig.~\ref{fig:lin.in.slides} where at first the ``faster'' mode (with small
negative amplitude) falls into the excised region,  while the ``slower''
mode (with positive amplitude) propagates in the same direction. We also
monitor the convergence factor for the error modes that have propagated 
outside the horizon, defined in terms of the $L_2$ norm over
the interval $x_H= 0.25\le x\le 1$ as
\begin{equation}
{\cal C}(x \ge x_H) = 
  \log_2 \left( \frac{ ||u_{\rho=1} ||_{2}}{ ||u_{\rho=2} ||_{2}} \right).
  \label{eq:convout}
\end{equation}
Second order accuracy would imply ${\cal C}(x \ge x_H)\ge 2$ since the analytic
solution vanishes outside the horizon.

\begin{figure}
\psfrag{ylabel}{${\cal C}(x \ge x_H)$}
\psfrag{xlabel}{time}
\psfrag{withdissip}{{\small with artificial dissipation}}
\psfrag{nodissip}{{\small with no artificial dissipation}}
  \includegraphics[width=8.5cm]{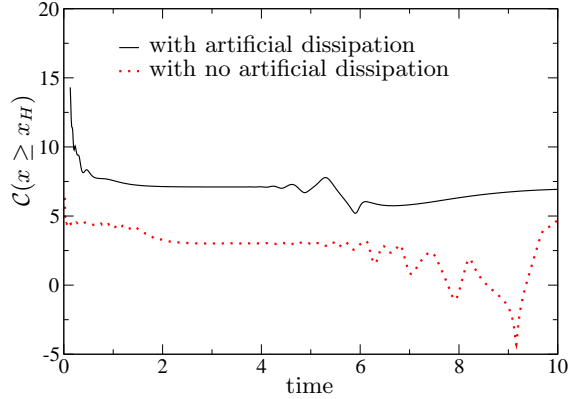}
\caption{Plot of the convergence factor ${\cal C}(x \ge x_H)$ as a  function
of time for the test run with an initial pulse inside the horizon. It is
evident that the addition of dissipation is necessary for long term convergence.}
\label{fig:lin.in.conv}
\end{figure}

In the region outside the horizon, the simulation consists purely of short
wavelength error. At first there is a high convergence rate but after about
6 crossing times the short wavelength error is unresolved on the grid.
The addition of dissipation to the evolution scheme greatly
extends the time for which convergence holds in the outer region, as shown in
Fig.~\ref{fig:lin.in.conv}. Dissipation is added by modifying (\ref{eq:fo})
according to
\begin{eqnarray}
  v_t\rightarrow  v_t + \epsilon_v D_+ D_- F D_+ D_- v
\label{eq:lin.test.singepulse}
\\
   u_t \rightarrow u_t +\epsilon_u D_+ D_- F D_+ D_- u,
\end{eqnarray}
where we set $\epsilon_v = \epsilon_u = 0.1 \cdot h^3$ and $F$ is a smooth
function with $F=b$ in the interior and $F = 0$ near both boundaries. 

\begin{figure}
\psfrag{xlabel}{$x$ axis}
\psfrag{ylabel}{$u$}
  \includegraphics[width=8.5cm]{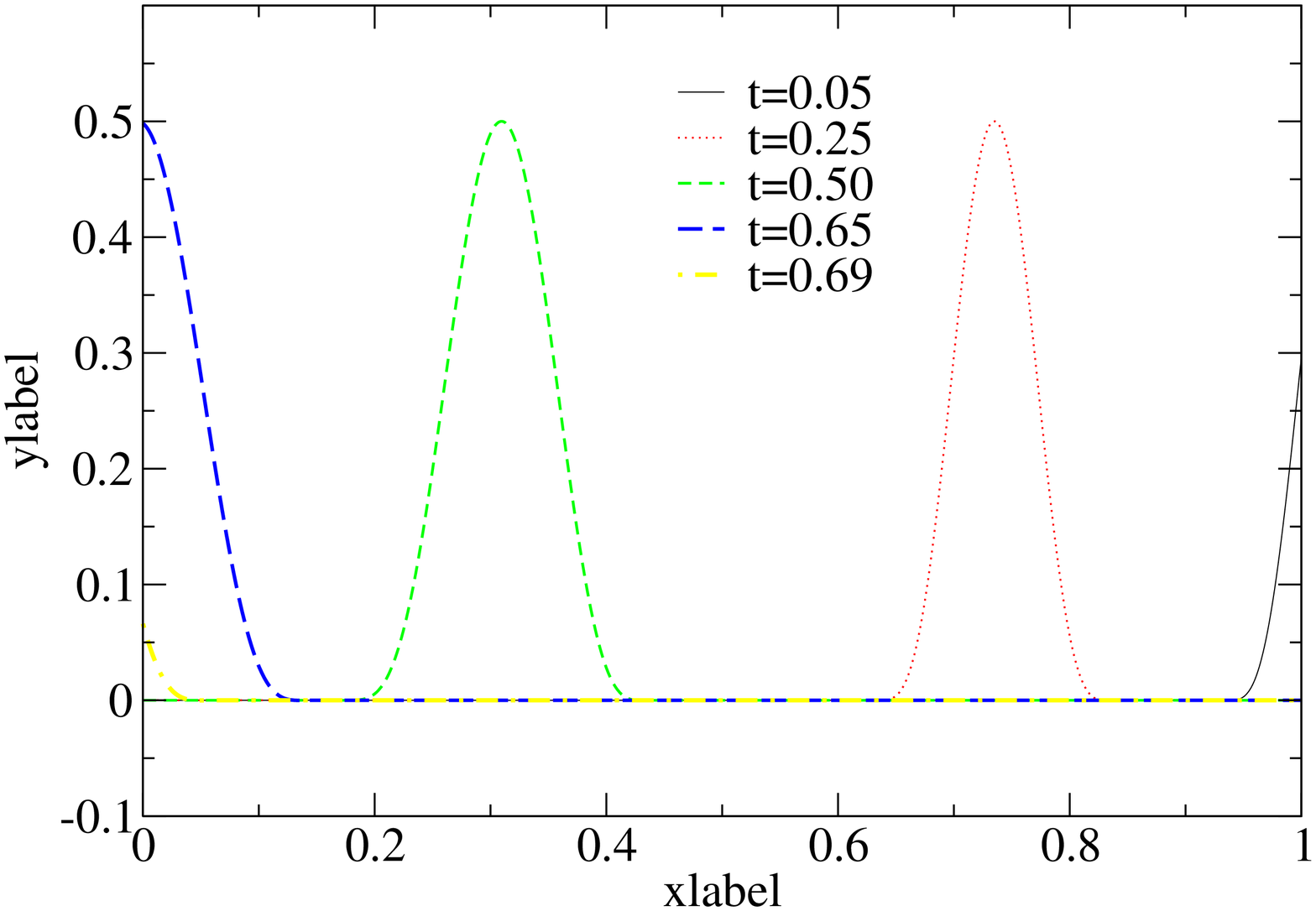}
  \includegraphics[width=8.5cm]{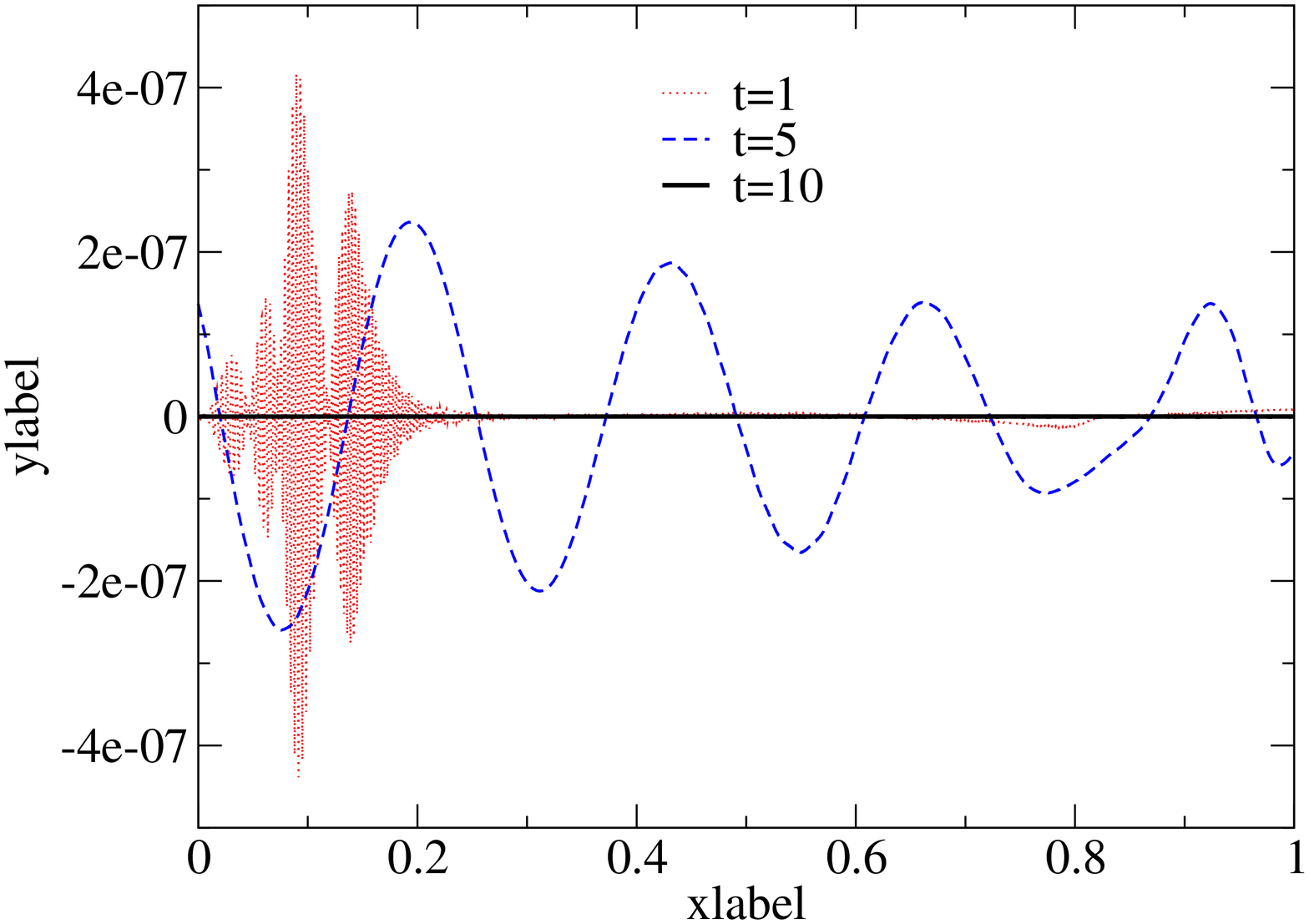}
\caption{Sequence of snapshots showing a pulse propagating across the grid.
The left graph shows the propagation of the main signal, while the right graph
illustrates the behavior of the residual error. The short wavelength error can
be eliminated by dissipation.}
\label{fig:lin.out.slides}
\end{figure}

In the second set of runs, the initial Cauchy data were set to zero,
$u(0,x)=u_t(0,x)=0$, and a wave was introduced through the outer boundary
$x=1$ by prescribing inhomogeneous Neumann data 
\begin{equation}
     g^{x\alpha}\partial_\alpha u|_{x=1}=
     (\frac{3}{4}u_x+\frac{1}{4}u_t)|_{x=1}=q(t).
     \label{eq:modneum}
\end{equation}
Figure~\ref{fig:lin.out.slides} shows snapshots of the
evolution for a wave generated by boundary data $q(t)$ consisting of a single
pulse. The pulse enters the outer boundary in the incoming mode of
(\ref{eq:lsol}) with characteristic velocity $-3/2$, propagates across the
blending region and leaves the grid at the excision boundary. The figure also
shows the remnant signal when no dissipation is added to the code. After the
initial pulse has entered the outer boundary, the Neumann data is homogeneous
and reflects any signal propagating to the right but signals propagating to the
left can leave the system across the excision boundary. At $t=1$, short
wavelength noise fills the region inside the horizon. By $t=5$, the short
wavelength error has either propagated through the excision boundary or has
been converted into longer wavelengths. At $t=10$, no visible signal remains.
The short wavelength noise at $t=1$ can be effectively eliminated by a small
amount of dissipation.

We checked long term global convergence by prescribing a periodic wave
entering though the outer boundary with the inhomogeneous Neumann data 
\begin{eqnarray}
   q(t) &=& 0, \quad \mbox{for} \quad t < 1/150
            \nonumber \\
   q(t) &=& \left. \Big( c \partial_x 
    + a \partial_t \Big) \Big[ A \sin^4\left(\pi [ x - 1.01 + 1.5 t ] 
        \right) \Big] \right|_{x=1}, \mbox{for}
       \quad t\ge 1/150.
\label{eq:lin.test.periodicpulse}
\end{eqnarray}
Again the initial Cauchy data was set to zero. The $C^3$ boundary data
(\ref{eq:lin.test.periodicpulse}) is initially set to zero to provide
$C^\infty$ consistency with the Cauchy data. No artificial
dissipation was added in this test.

\begin{figure}
\psfrag{xlabel}{time}
\psfrag{ylabel}{${\cal C}$}
  \includegraphics[width=8.5cm]{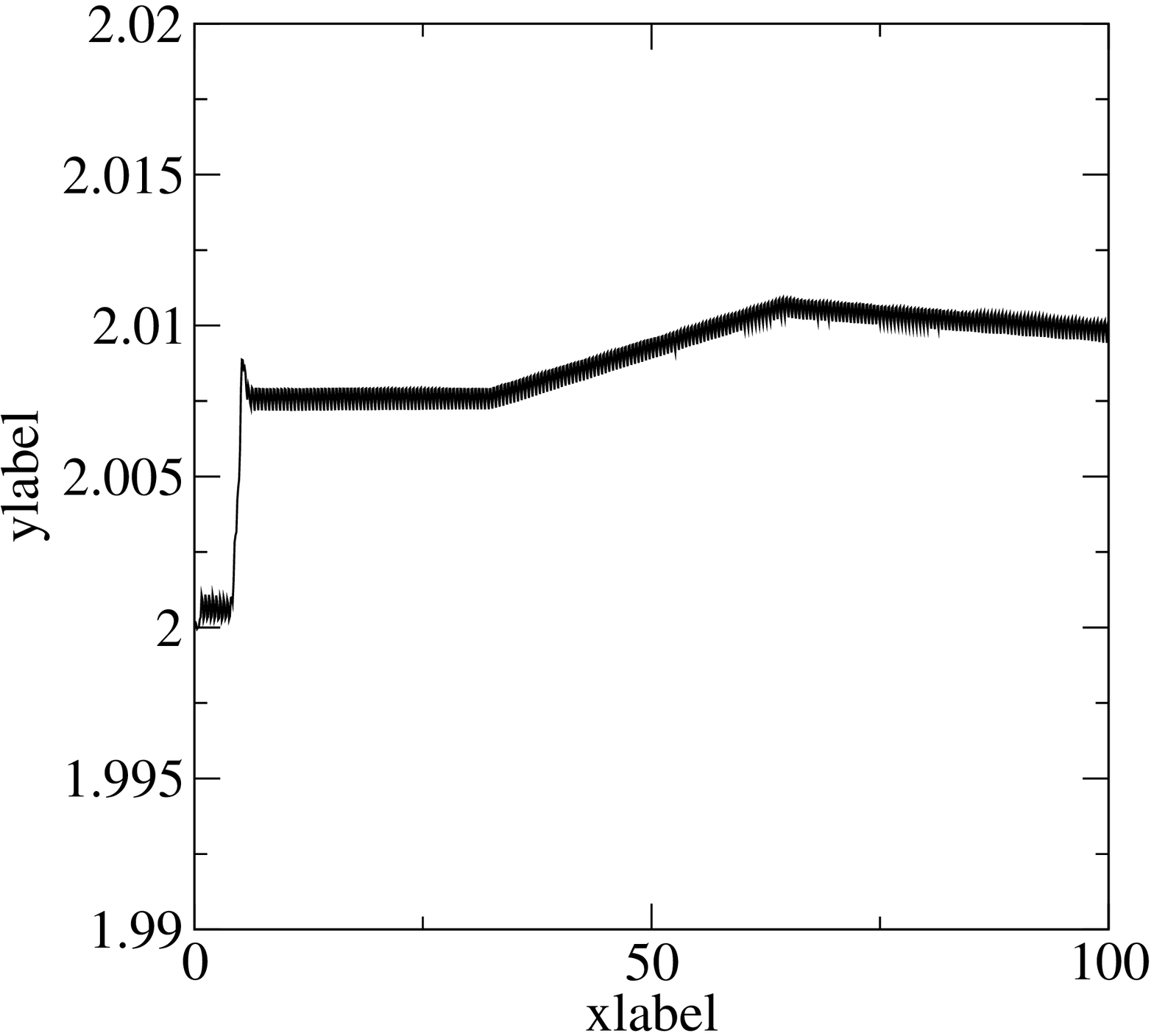}
\psfrag{xlabel}{$x$ axis}
\psfrag{ylabel}{Error in $u$}
\psfrag{42}[lb]{$\;u_{\rho=4}-u_{\rho=2}$}
\psfrag{84}[l]{$(u_{\rho=8}-u_{\rho=4}) \times 4$}
  \includegraphics[width=8.5cm]{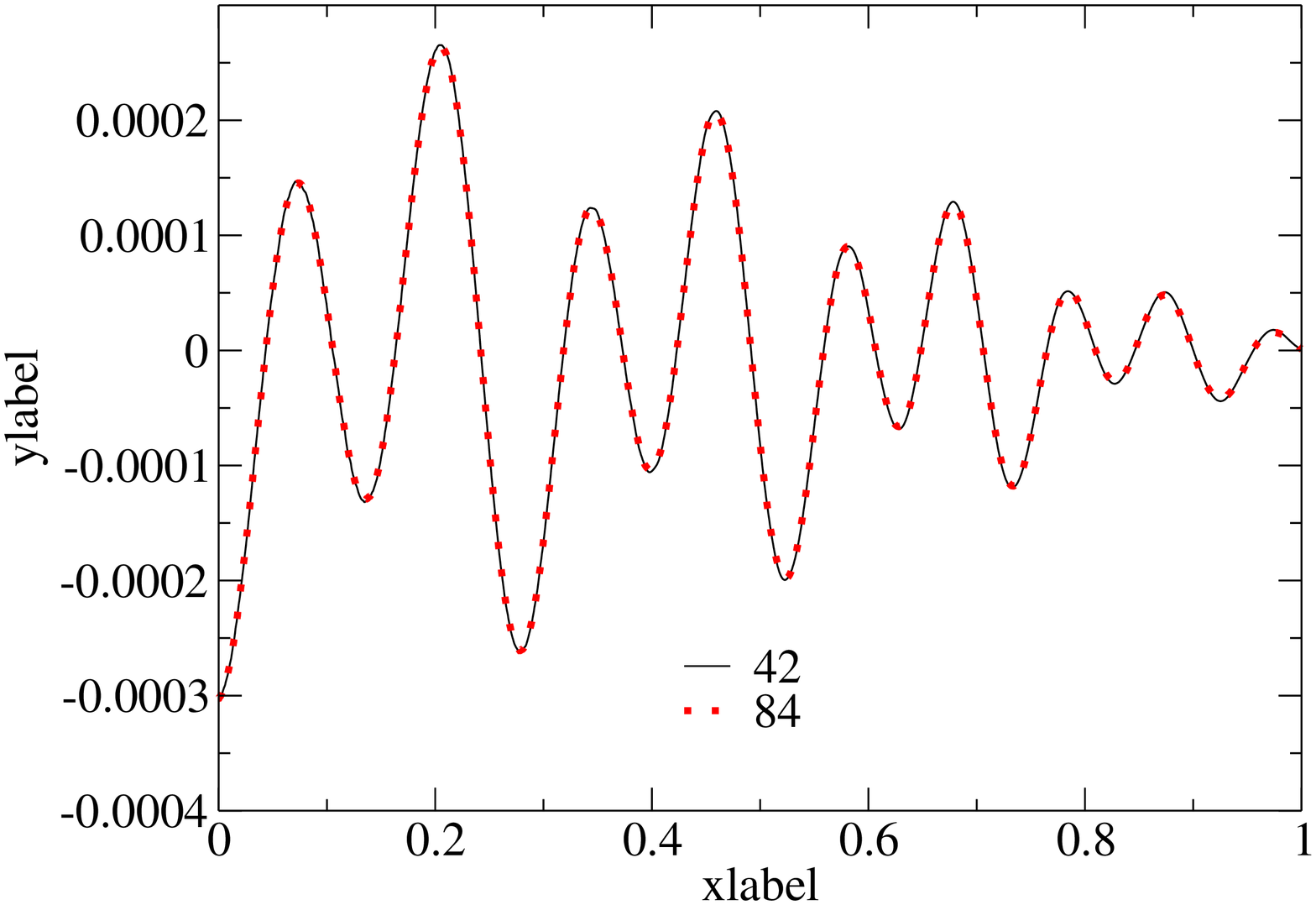}
\caption{Convergence plots for the test run with a periodic wave
introduced through the outer (timelike) boundary. On the left, the plot of
the convergence factor ${\cal C}$  vs. time indicates good second order
convergence up to the end of the simulation at $t=100$. The quality of the
long term performance is reinforced by the right graph which shows that,
at the end of the simulation, the rescaled error profiles are in nearly
perfect agreement. No artificial dissipation was introduced.}
\label{fig:lin.long}
\end{figure}

We measured Cauchy convergence of the numerical solution by
monitoring the convergence factor
\begin{equation}
      {\cal C} = \log_2 \left( \frac{ ||u_{\rho=1} 
    - u_{\rho=2} ||_2}{ ||u_{\rho=2} - u_{\rho=4} ||_2 } \right).
    \label{eq:conv}
\end{equation}
For the given data and grid sizes, the code displays second order accuracy
to within less than one percent. Figure~\ref{fig:lin.long} plots  the
convergence factor as well as a snapshot of the residual finite difference
error at the end of the simulation, at $t=100$.

\section{The quasi-linear excision problem}
\label{sec:qlinexcis}

\subsection{Algorithms for the quasi-linear wave equation}
\label{sec:qlinweq}

We extend our algorithms to the quasi-linear wave equation (\ref{eq:nlwave})
in such a way that the outer algorithm continues to obey a
semi-discrete monopole conservation law in the case of homogeneous Neumann
boundary conditions.
We again set $c=b-a^2$ and treat the case of
time-dependent, non-constant coefficients.
We generalize the outer algorithm (\ref{eq:onclinwave}) to the
finite-difference form
\begin{equation}
  W:= \partial_t \bigg(\frac{1}{\Phi}\Phi_t \bigg)
         -\partial_t\bigg(\frac{a}{\Phi}D_0 \Phi\bigg)
	 -D_0\bigg(\frac {a}{\Phi} \Phi_t\bigg)
       -\frac{1}{2} D_-\bigg( (A_+ \frac{c}{\Phi}) D_+\Phi \bigg)
         -\frac{1}{2} D_+\bigg( (A_- \frac{c}{\Phi}) D_- \Phi\bigg) =0, 
   \label{eq:oncnlinwave}
\end{equation}
and generalize the horizon algorithm (\ref{eq:nchlinwave}) to 
\begin{equation}
   V:= \partial_t \bigg(\frac{1}{\Phi}\Phi_t \bigg)
         -\partial_t\bigg(\frac{a}{\Phi}D_0 \Phi\bigg)
	 -D_0\bigg(\frac {a}{\Phi} \Phi_t\bigg) 
	  -\frac{1}{2} D_-\bigg( (A_+ \frac{b}{\Phi}) D_+\Phi \bigg)
         -\frac{1}{2} D_+\bigg( (A_- \frac{b}{\Phi}) D_- \Phi\bigg)
           +D_0\bigg(\frac{a^2}{\Phi} D_0 \Phi \bigg) =0 . 
   \label{eq:nchnlinwave}
\end{equation}
We blend these algorithms using the quasi-linear version of
(\ref{eq:blend}), i.e $B=0$ with
\begin{equation}
   B= W - D_-\bigg( (A_+ \frac {f a^2}{\Phi}) D_+ \Phi \bigg)
          + D_0\bigg(\frac {fa^2}{\Phi} D_0 \Phi\bigg)
     = W +\frac{h^2}{4} D_+ D_- \bigg(\frac{fa^2}{\Phi}D_+ D_- \Phi\bigg) .
   \label{eq:nlblend}
\end{equation}

As the boundary condition for the horizon algorithm at the spacelike excision
boundary, we use the extrapolation conditions (\ref{eq:uextrap}), now written as
\begin{equation}
   h^3 D_+^3 \Phi_0=0,\quad h^3 D_+^3 \Phi_1=0. 
\label{eq:extrap}
\end{equation} 
Similarly, as homogeneous Dirichlet condition for the outer algorithm at a
timelike outer boundary we retain the linear form $\Phi_t=0$. We formulate the
homogeneous Neumann condition for the outer algorithm to establish
semi-discrete monopole conservation corresponding to the continuum conservation
law
\begin{equation}
     \partial_t \int_0^1 (\frac{\Phi_t}{\Phi}-\frac{a\Phi_x}{\Phi})dx =0.
\end{equation}
We proceed as follows.

For periodic boundary conditions, (\ref{eq:oncnlinwave}) implies the
semi-discrete conservation law $\partial_t {\cal Q}=0$ where
\begin{equation} 
    {\cal Q}= h\sum_\nu \bigg (\frac {\Phi_{t\nu}}{\Phi_\nu}
             - \frac {a_\nu D_0\Phi_\nu}{\Phi_\nu} \bigg). 
\label{eq:qper}
\end{equation}
This conservation law can be extended to a homogeneous Neumann
boundary condition by adding boundary terms to (\ref{eq:qper}) in the form
\begin{eqnarray} 
     {\cal Q}&=& h\sum_{\nu=1}^{N-1}\bigg (\frac {\Phi_{t\nu}}{\Phi_\nu}
             - \frac {a_\nu D_0\Phi_\nu}{\Phi_\nu}\bigg)
                           \nonumber \\
          &+&\frac{h}{2}\bigg (\frac {\Phi_{t0}}{\Phi_0}
             - \frac {a_0 D_+\Phi_0}{\Phi_0}
             +\frac {\Phi_{tN}}{\Phi_N}
             - \frac {a_N D_-\Phi_N}{\Phi_N} \bigg).
\end{eqnarray}
We generalize the Neumann condition (\ref{eq:homneum}) to the
non-linear form
\begin{equation}
 {\cal N}:= \frac{1}{2}\bigg( (A_+ \frac {c}{\Phi})D_+ \Phi
      +(A_- \frac{c}{\Phi})D_- \Phi
  +  (A_+ \frac{a}{\Phi})T_+\Phi_t+ (A_- \frac{a}{\Phi})T_- \Phi_t \bigg)
  +\frac{h^2}{4} \partial_t \bigg(\frac{a}{\Phi}\bigg)D_+ D_- \Phi=0.
  \label{eq:nlhomneum}
\end{equation}
As a result of the nonlinearities, (\ref{eq:nctimneumN}) and (\ref{eq:nctimneum0})
are modified to  
\begin{eqnarray}
  \frac{h}{2}W_N+{\cal N}_N:&=&\frac{h}{2}\bigg(
               \partial_t( \frac {\Phi_{tN}}{\Phi_N})
              -\partial_t ( \frac {a_N}{\Phi_N})D_- \Phi_N \bigg)
	        +(A_- \frac{c_N}{\Phi_N}) D_-\Phi_N+ 
		(A_- \frac {a_N}{\Phi_N})\Phi_{t(N-1)}
   \label{eq:nctnlwneumN}   \\    
  -\frac{h}{2}W_0+{\cal N}_0:&=&-\frac{h}{2}\bigg(
               \partial_t( \frac {\Phi_{t0}}{\Phi_0})
              -\partial_t ( \frac {a_0}{\Phi_0})D_+ \Phi_0 \bigg)
	        +(A_+ \frac{c_0}{\Phi_0}) D_+\Phi_0+ 
		(A_+ \frac {a_0}{\Phi_0})\Phi_{t1}.
   \label{eq:nctnlwneum0}    
\end{eqnarray}
Setting $W_\nu=0$ for $1\le\nu\le N-1$, a straightforward calculation
leads to the nonlinear version of (\ref{eq:fluxcons}), i.e.
\begin{equation}
     {\cal Q}_t =\frac{h}{2}W_N+{\cal N}_N +\frac{h}{2}W_0-{\cal N}_0.
\end{equation}
We implement the homogeneous Neumann condition in the form
\begin{eqnarray}
  {\cal N}_N&=&\frac{h}{2}\bigg( \partial_t( \frac {\Phi_{tN}}{\Phi_N})
              -\partial_t ( \frac {a_N}{\Phi_N})D_- \Phi_N \bigg)
	        +(A_- \frac{c_N}{\Phi_N}) D_-\Phi_N+ 
		(A_- \frac {a_N}{\Phi_N})\Phi_{t(N-1)}=0
   \label{eq:nctnlneumN}   \\    
  {\cal N}_0&=&-\frac{h}{2}\bigg( \partial_t( \frac {\Phi_{t0}}{\Phi_0})
              -\partial_t ( \frac {a_0}{\Phi_0})D_+ \Phi_0 \bigg)
	        +(A_+ \frac{c_0}{\Phi_0}) D_+\Phi_0+ 
		(A_+ \frac {a_0}{\Phi_0})\Phi_{t1}=0.
   \label{eq:nctnlneum0}    
\end{eqnarray}
Then the requirement that $W_N=W_0=0$ leads to the semi-discrete monopole
conservation ${\cal Q}_t=0$.

\subsection{Tests of the quasi-linear excision algorithm}
\label{sec:qltests}

The excision tests performed with the quasi--linear code were based upon the
ones performed with the linear code, with the (time independent) coefficients
of the wave operator having spatial dependence determined by
(\ref{eq:exc_background}) and with the same Cauchy data or boundary data
now prescribed for $(\Phi-1)$ as were prescribed for the linearized field
$u$ in Sec.~\ref{sec:ltests}.

\begin{figure}
\psfrag{xlabel}{$x$ axis}
\psfrag{ylabel}{$\Phi$}
  \includegraphics[width=8cm]{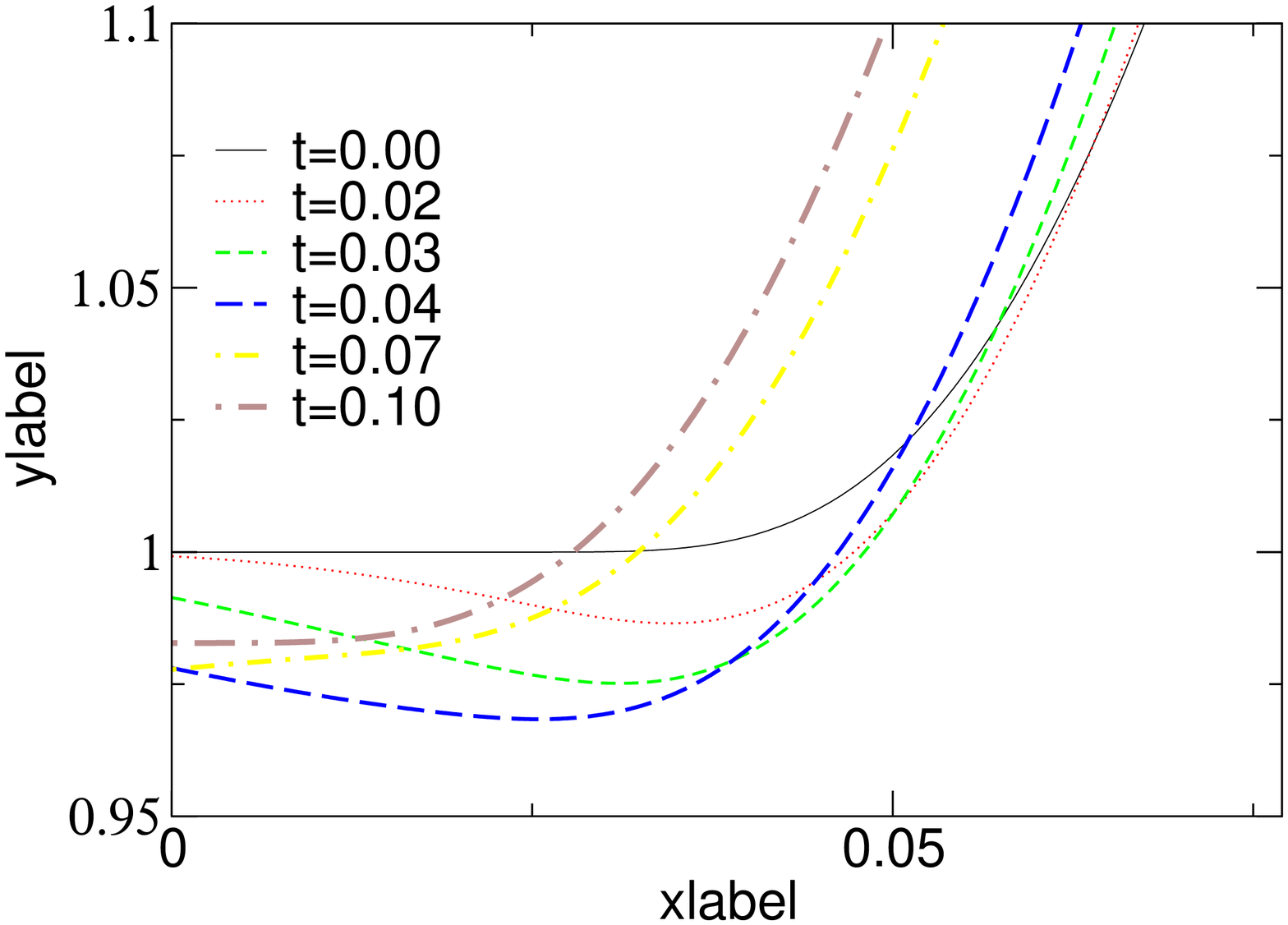}
  \includegraphics[width=8cm]{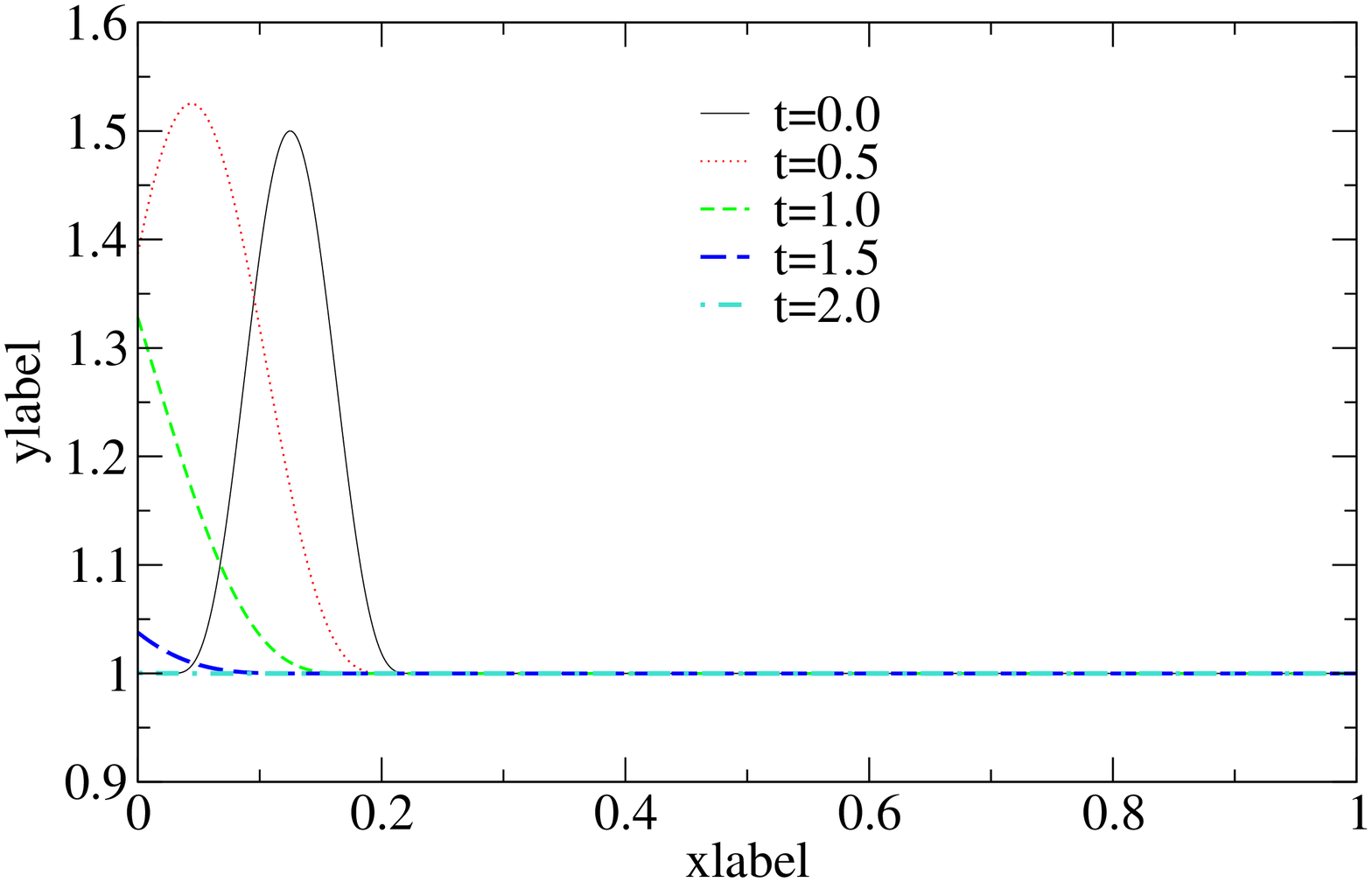}
\caption{Sequence of snapshots showing the initial pulse falling into the
excised region, evolved with the quasi-linear system. The left graph shows the
faster mode (with negative amplitude), while the right graph shows the slower
mode making its way through the excision boundary. Grid-resolution for this
simulation was $h=1/8000$.}
\label{fig:nonlin.in.slides}
\end{figure}

\begin{figure}
\psfrag{ylabel}{${\cal C}(x \ge x_H)$}
\psfrag{xlabel}{time}
\psfrag{withdissip}{{\small with artificial dissipation}}
\psfrag{nodissip}{{\small with no artificial dissipation}}
  \includegraphics[width=8.5cm]{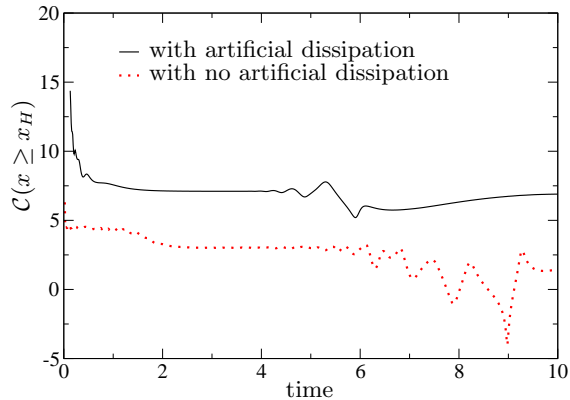}
\caption{Plot of the convergence factor ${\cal C}(x \ge x_H)$ vs. time.
for the non-linear test run for an initial pulse inside the horizon.
Addition of dissipation is necessary for long term convergence.}
\label{fig:nonlin.in.conv}
\end{figure}

The initial data for the first set of runs consisted of a pulse of compact
support inside the horizon with homogeneous Dirichlet data at the outer
boundary. The initial peak amplitude of the pulse (\ref{eq:cpulse}) was
$\Phi_{max}=1.5$, putting it in the nonlinear regime. Since the characteristic
speeds determined by the background metric are the same as in the linearized
case, the snapshots of the quasi-linear evolution shown in
Fig.~\ref{fig:nonlin.in.slides} are qualitatively similar to the linearized
ones in Fig.~\ref{fig:lin.in.slides}. There is no visible propagation of the
signal into the region outside the horizon. The convergence factor ${\cal C}(x
\ge x_H)$ for the error modes that have propagated outside the horizon is
defined  by replacing $u$ by $(\Phi-1)$ in (\ref{eq:convout}). As in the linear
case, the short wavelength error in the exterior region is unresolved on the
grid. The plots in Fig.~\ref{fig:nonlin.in.conv} show that dissipation must be
added to the code to obtain long term convergence.

\begin{figure}
\psfrag{xlabel}{$x$ axis}
\psfrag{ylabel}{$\Phi$}
  \includegraphics[width=8.5cm]{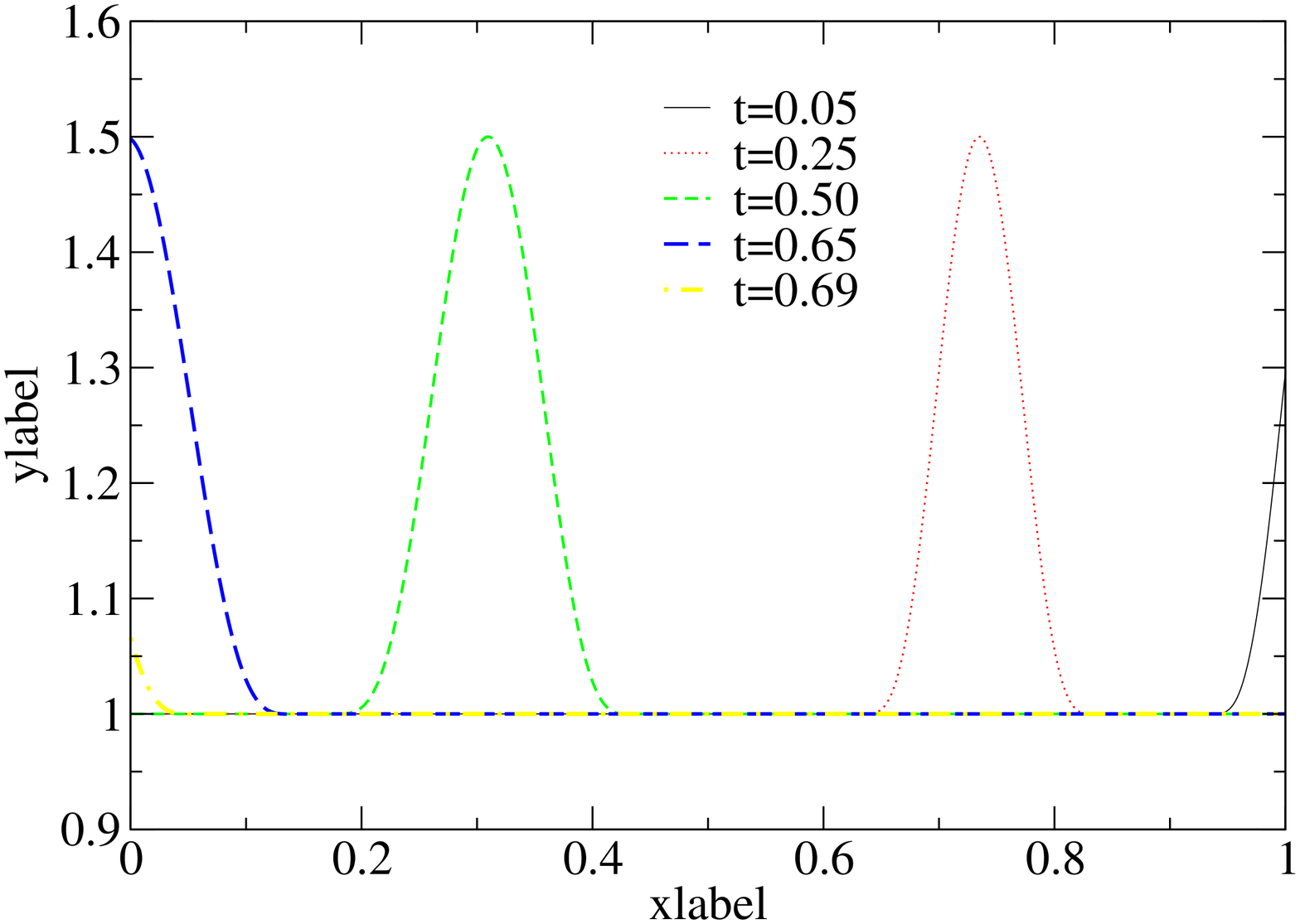}
\psfrag{ylabel}{$(\Phi-1)$}
  \includegraphics[width=8.5cm]{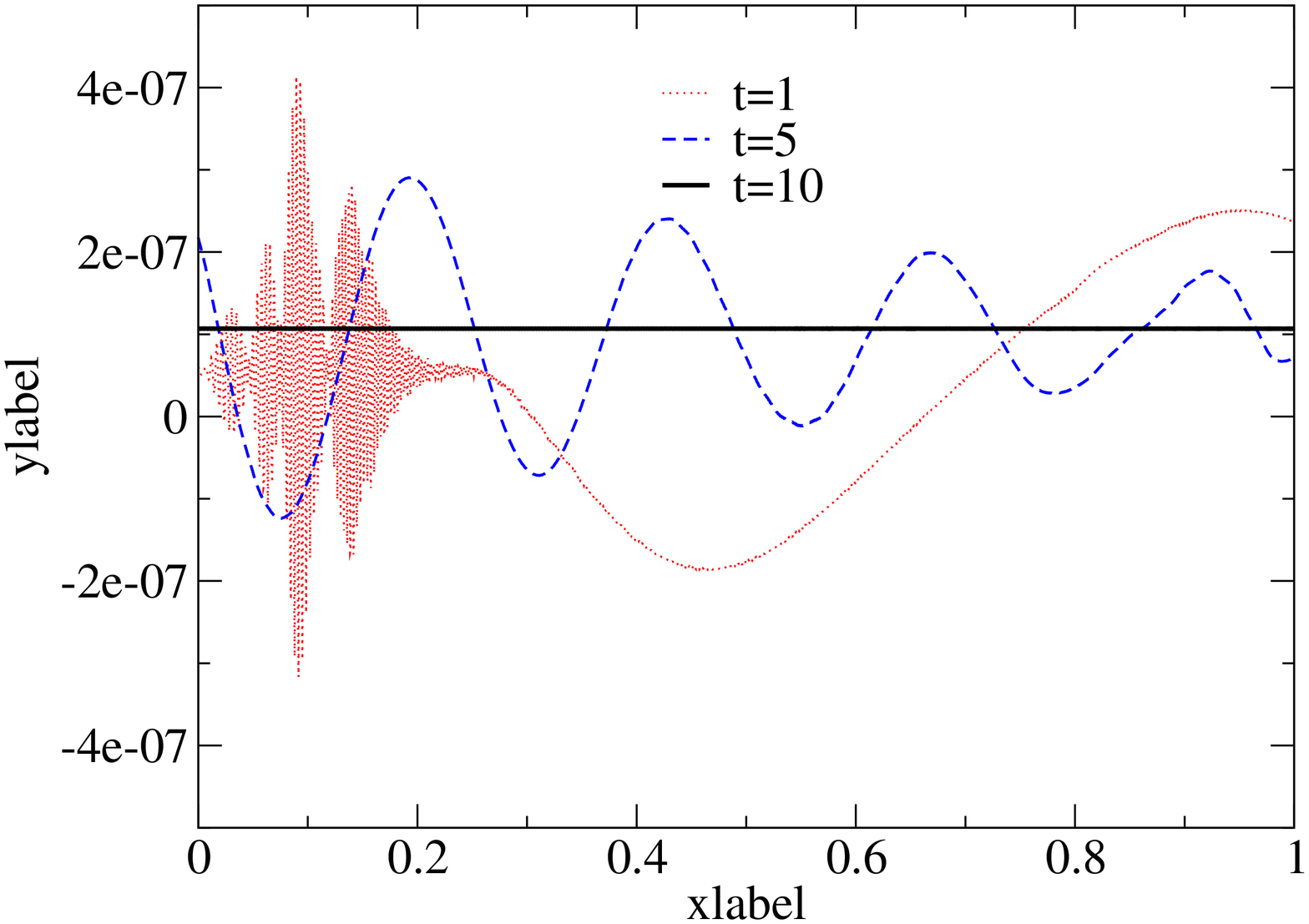}
\caption{Sequence of snapshots showing a pulse propagating across the grid
with resolution $h=1/2000$. The left graph shows the propagation of the
signal itself, while the right graph illustrates the behavior of the
residual modes. The long wavelength mode, which is apparent at $t=1$, decays
by $t=5$ and no visible signal remains at $t=10$.}
\label{fig:nonlin.out.slides}
\end{figure}

In the second set of non-linear runs, the initial Cauchy data were set to
zero and a wave introduced through the outer boundary by prescribing
inhomogeneous Neumann data.  Figure~\ref{fig:nonlin.out.slides} shows
snapshots of the evolution for a wave consisting of a single pulse 
entering the outer boundary. No artificial dissipation
was used. The main signal behaves qualitatively
similar to the linearized case in Figure~\ref{fig:lin.out.slides} but the
error modes behave differently. At $t=1$ the short wavelength error again fills
the region inside
the horizon but now long wavelength error has also been excited
which extends to the outer
boundary. This corresponds to mixing in the exponential modes
\begin{equation}
         \Phi=e^{\lambda(t-\frac{2x}{3})} ,
	 \label{eq:expmode}
\end{equation}
which are exact solutions (for any $\lambda$) to the quasi-linear wave
equation (\ref{eq:nlwave}) in the region $x\ge 3/4$. These modes are
consistent with the homogeneous Neumann boundary condition (\ref{eq:modneum})
in effect after the pulse has entered the system and $q(t)$ vanishes.
However, the mode (\ref{eq:expmode}) is not excited with positive $\lambda$
and decays by $t=5$. No visible signal remains at $t=10$. Thus the discrete
monopole conservation built into the quasi-linear system
suppress exponentially growing long wavelength modes.

\begin{figure}
\psfrag{xlabel}{time}
\psfrag{ylabel}{${\cal C}$}
  \includegraphics[width=8.5cm]{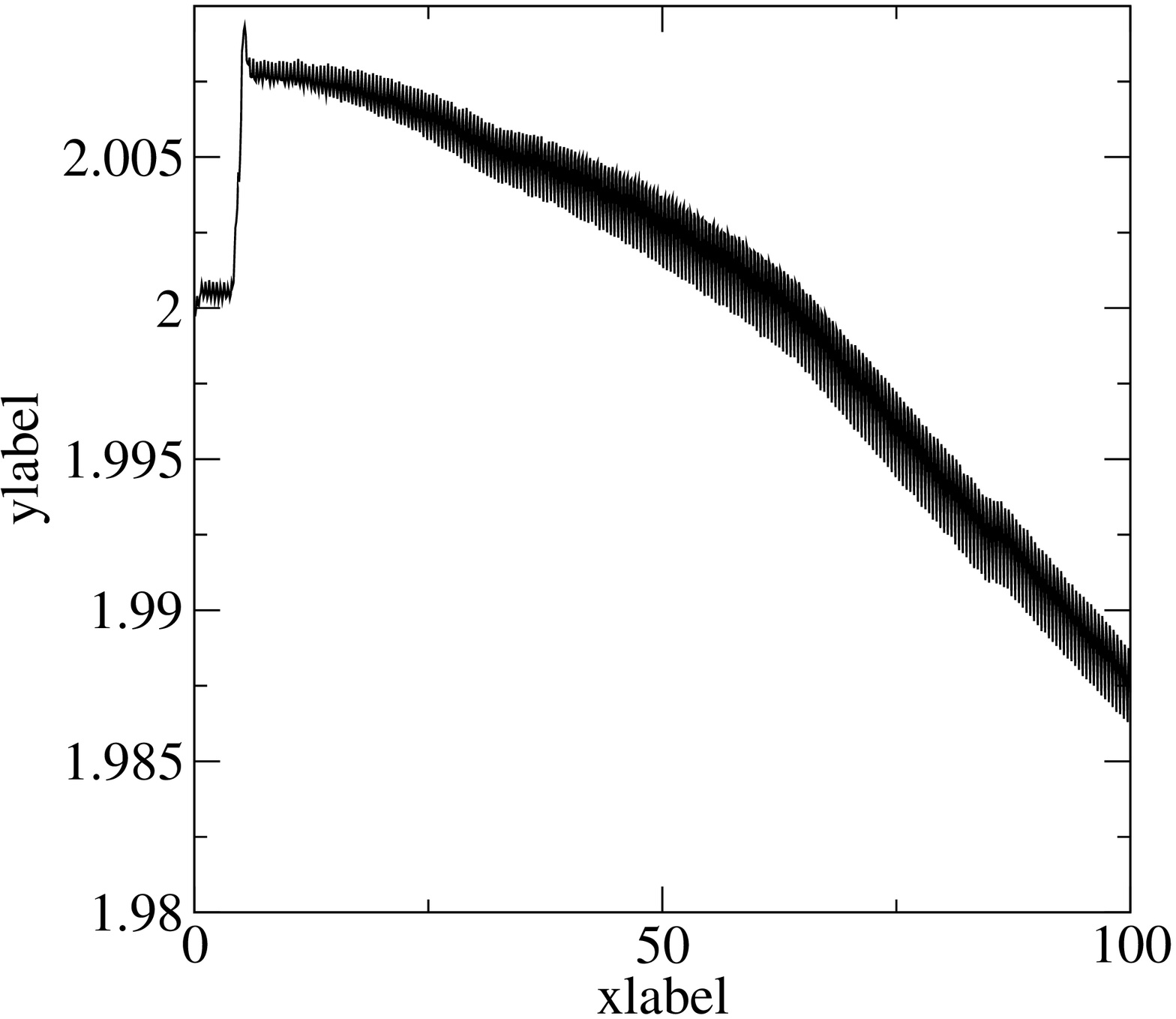}
\psfrag{xlabel}{$x$ axis}
\psfrag{ylabel}{Error in $\Phi$}
\psfrag{42}[lb]{$\;\Phi_{\rho=4}-\Phi_{\rho=2}$}
\psfrag{84}[l]{$(\Phi_{\rho=8}-\Phi_{\rho=4}) \times 4$}
  \includegraphics[width=8.5cm]{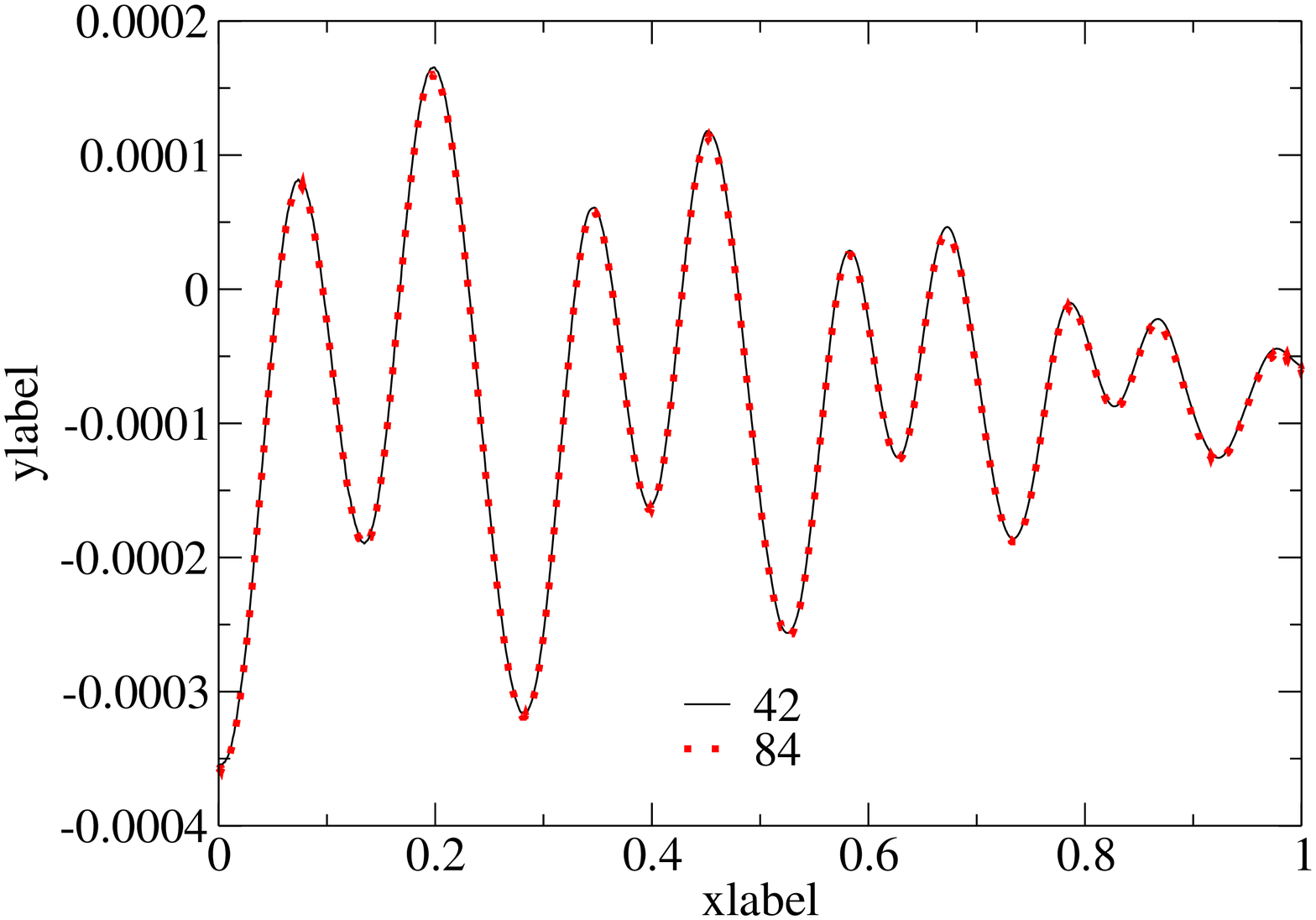}
\caption{Convergence plots for the non-linear test run with a periodic pulse
introduced through the outer (timelike) boundary. On the left, the time
dependence of the convergence factor ${\cal C}$ indicates good second order
convergence. This result is reinforced by the right graph where, at the end
of the simulation at $t=100$, the rescaled error profiles show a nearly
perfect agreement. No artificial dissipation was used in this test.}
\label{fig:nonlin.long}
\end{figure}

A further test of long term performance is the introduction of a periodic
wave through the outer boundary by prescribing the inhomogeneous Neumann
data (\ref{eq:lin.test.periodicpulse}). We measured Cauchy convergence by
monitoring the convergence factor ${\cal C}$ (\ref{eq:conv}) applied to the
numerical solution for $\Phi$. Figure~\ref{fig:nonlin.long} plots ${\cal
C}$ as well as a snapshot of the residual finite difference error at
$t=100$, when the simulation was ended. The code displays second order
accuracy to better than 1 percent. No artificial dissipation was used
in the test.

\section{Discussion}
\label{sec:disc}

We have based successful simulations of the model excision problem for a
1-dimensional quasi-linear wave equation on evolution and boundary algorithms
which were proved to be stable in the linear case. The quasi-linear algorithm
incorporates a conserved scalar monopole moment which  suppresses the
excitation of long wavelength exponentially growing modes, such as
(\ref{eq:expmode}), which are latent in the system when Neumann boundary
conditions are used. The global excision strategy involves matching an exterior
evolution algorithm to a horizon algorithm for evolution of the interior
region. The exterior algorithm admits simple implementation of an outer
boundary condition and it obeys a discrete version of flux-energy conservation
which guarantees stability. Inside the horizon, this energy is not
positive-definite and the exterior algorithm is unstable. The horizon algorithm
is stable in the interior region and admits a stable extrapolation algorithm to
update the spacelike excision boundary. The choice of horizon algorithm
presented here is based upon centered differencing. Other choices are possible
in the model 1-dimensional problem. In particular, a horizon algorithm based
completely upon one-sided differencing is stable and requires no extrapolation
at the inner boundary. This would be a very attractive feature if it could be
extended to the higher dimensional problem. 

Preliminary investigations indicate that the computational algorithms presented
here can be taken over in a fairly straightforward manner to the harmonic
formulation of the full 3-dimensional gravitational problem. In that case the
Einstein equations reduce to to the quasi-linear form
\begin{equation}
  \frac{1}{\sqrt{-g}}\partial_\alpha
        (\sqrt{-g}g^{\alpha\beta}\partial_\beta g^{\mu\nu})=S^{\mu\nu} ,
	\label{eq:he}
\end{equation}
where $S^{\mu\nu}$ consists of lower differential order terms which do
not contribute to the principle part. These equations have a well-posed
initial-boundary value problem when the boundary data is explicitly
prescribed~\cite{szi03}. 

The principle part of (\ref{eq:he}) can be treated in the same way as a
quasi-linear wave equation (\ref{eq:wave}) for each component of the
metric. From the point of view of designing a code, the two main
differences from our model excision problem are (i) the  the 3-dimensional
nature of the full problem and (ii) the constraints which enter into the
boundary conditions. 

The evolution algorithms readily extend to 3-dimensions. When $S^{\mu\nu}$ is
neglected, the flux conservative form of (\ref{eq:he}) can be carried over to
the 3 dimensional discretized system to obtain discrete analogues of the
scalar wave monopole conservation. There results a conserved quantity representing
a spatially averaged rate of growth for each metric component, which has
monopole symmetry for the $g^{tt}$ component, dipole symmetry for the $g^{ti}$
components and quadrupole symmetry for the $g^{ij}$ components. The exact form
of the conserved quantities depends upon how the principle part is split off in
conservative form, with (\ref{eq:he}) just one possibility. See the discussion
in~\cite{badh} for how a splitting analogous to the logarithmic form
(\ref{eq:nlwave}) for the quasi-linear scalar problem leads to accurate long
term simulation of the gauge wave problem; it remains to be seen if this
splitting is effective in more general spacetimes. When $S^{\mu\nu}$ is neglected and the
metric coefficients in the wave operator are frozen,
semi-discrete energy conservation also results for each metric component in the
case of periodic boundary conditions. With a cubic grid boundary, the results
for the quasi-linear wave equation generalize to dissipative Dirichlet and
Neumann boundary conditions for the metric components. The chief
complication is the application of Neumann conditions at the edges and corners
of the cube. Here among possible approaches are the use of the {\em summation
by parts} technique, as in~\cite{lsusbp}, or the introduction of a smooth
boundary which is treated by interpolation, as in~\cite{krdir,krneum}. 

The constraints on the boundary data present another complication. The
constraints on the system (\ref{eq:he}), which guarantee that the solutions
satisfy Einstein's equations, are the harmonic coordinate conditions
\begin{equation}
               \partial_\nu(\sqrt{-g}g^{\mu\nu})=\hat H^\mu
\end{equation}
where $\hat H^\mu$ are explicit harmonic driving functions. These constraints
are satisfied by the solutions of (\ref{eq:he}) if a certain mixture of
Dirichlet and Neumann boundary conditions are used for the metric
components~\cite{szi03}. In that case, the constraints are satisfied when
homogeneous boundary data are given and the initial-boundary problem for
Einstein's equations is well-posed. In the case of inhomogeneous
constraint-preserving boundary data it is not known whether the system
remains well-posed.

It should be emphasized that, in addition to the computational difficulty,
there  are analytic and geometric problems in treating black holes. There is
the possibility of exponentially growing perturbations to the analytic
problem in the inner region near the excision boundary.  There is the
nonlocal nature of the {\em true} (null) event horizon. Although the
spacelike apparent horizon can be traced out as the evolution proceeds, it is
impossible to locate the null event horizon until the exterior evolution is
complete. The matching of an inner horizon algorithm to an outer algorithm
must be carried out across an artificial horizon which results from a given
choice of shift, as in our model problem. In the black hole excision problem,
such an artificial horizon could be defined in terms of the hypersurface
across which the evolution goes from superluminal to subluminal.

Pretorius~\cite{pret} has recently obtained promising results using a second
order harmonic code to simulate a black hole by means of excision. We are now
in the process of applying the new techniques presented here to this problem.

\begin{acknowledgments}

We thank M. Babiuc, L. Lehner and M. Tiglio for their input. This work was
supported by the National Science Foundation under grant PH-0244673 to the
University of Pittsburgh.

\end{acknowledgments}


\begin{thebibliography}{40}


\bibitem{Garf} D. Garfinkle, {\em Phys.Rev.}, {\bf D65}, 044029 (2002).

\bibitem{szi03} B. Szil\'{a}gyi and J. Winicour, {\em Phys. Rev.}, {\bf D68},
041501 (2003).

\bibitem{pret} F. Pretorius, {\em Class.Quant.Grav.} {\bf 22}, 425 (2005).

\bibitem{adm}  R. Arnowitt, S. Deser and C. Misner, in {\it Gravitation: An
Introduction to Current Research}, ed. L. Witten (New York, Wiley, 1962).

\bibitem{lsulsw} G. Calabrese, L. Lehner, D. Neilsen, J. Pullin, O. Reula, O.
Sarbach, and M. Tiglio, {\em Class. Quant. Grav.} {\bf 20}, L245, (2003).

\bibitem{calab} G. Calabrese, {\em Phys.Rev. D} {\bf 71}, 027501 (2005). 

\bibitem{apples} www.appleswithapples.org

\bibitem{mex1m} M. Alcubierre, {\em et al}, {\em Class. Quantum Grav.},
{\bf 21}, 589 (2004).

\bibitem{badh} ``Some mathematical problems in numerical relativity'',
M. Babiuc, B. Szil\'{a}gyi and J. Winicour, gr-qc/0404092.

\bibitem{lsugw} M. Tiglio,
L. Lehner, D. Neilsen, {\em Phys.Rev. D} {\bf 70}, 104018 (2004).

\bibitem{krsbp} H.-O. Kreiss and G. Scherer, {\em SIAM J. Numer. Anal.}
{\bf 29}, 640 (1992).

\bibitem{lsusbp} G. Calabrese, L. Lehner, O. Reula, O. Sarbach,
and M. Tiglio, {\em Class.Quant.Grav.} {\bf 21}, 5735 (2004).

\bibitem{lsude} L. Lehner, D. Neilsen, O. Reula and
M. Tiglio,  {\em Class.Quant.Grav.} {\bf 21}, 5819 (2004).

\bibitem{krsecor} H.-O. Kreiss, N.~A. Peterson and J. Ystr\"{o}m,
{\em SIAM J. Numer. Anal.} {\bf 40}, 1940 (2002).

\bibitem{kreissort} H.-O. Kreiss and O.~E. Ortiz,
{\em Lect. Notes Phys.}, {\bf 604}, 359 (2002).

\bibitem{krdir} ``A second order accurate embedded boundary method
for the wave equation with Dirichlet data'', H.-O. Kreiss and N.~A. Peterson,
preprint UCRL-JRNL-202686, Lawrence Livermore National Lab (2004).

\bibitem{krneum} ``Difference approximations of the Neumann problem for the
second order wave equation'', H.-O. Kreiss, N.~A. Peterson and J. Ystr\"{o}m,
preprint UCRL-JC-153184, Lawrence Livermore National Lab (2003).

\bibitem{alcsch} M. Alcubierre and B. Schutz, {\em J. Comput. Phys.}, {\bf 112},
44 (1994).

\bibitem{sundstrom} B. Gustafsson, H.-O. Kreiss and A. Sundstrom,
{\em Mathematics of Computation}, {\bf26}, 649 (1972).

\bibitem{kreiss} B. Gustafsson, H.-O. Kreiss and J. Oliger, {\em Time
Dependent Problems and Difference Methods} (Wiley, NY, 1995).

\end{thebibliography}
\end{document}